\pdfoutput=1
\RequirePackage{ifpdf}%

\documentclass[journalnocopy]{vgtc}

\nocopyrightspace

\usepackage{mathptmx}
\usepackage{graphicx}
\usepackage{times}
\usepackage{subfigure}
\usepackage{url}

\onlineid{0}

\vgtccategory{Research}

\vgtcinsertpkg

\title{A Moving Least Squares Based Approach for\\
Contour Visualization of Multi-Dimensional Data}

\author{Chris W. Muelder, Nick Leaf, Carmen Sigovan, \& Kwan-Liu Ma }
\authorfooter{
\item Authors are with VIDi group at U. C. Davis, e-mail: muelder/ma@cs.ucdavis.edu, njleaf/cmsigovan@ucdavis.edu.
}

\shortauthortitle{Muelder \MakeLowercase{\textit{et al.}}: A Moving Least Squares Based Approach for Countour Visualization of Multi-Dimensional Data}

\abstract{
Analysis of high dimensional data is a common task.  Often, small multiples are used to visualize 1 or 2 dimensions at a time, such as in a scatterplot matrix. Associating data points between different views can be difficult though, as the points are not fixed.  Other times, dimensional reduction techniques are employed to summarize the whole dataset in one image, but individual dimensions are lost in this view.  In this paper, we present a means of augmenting a dimensional reduction plot with isocontours to reintroduce the original dimensions.  By applying this to each dimension in the original data, we create multiple views where the points are consistent, which facilitates their comparison.  Our approach employs a combination of a novel, graph-based projection technique with a GPU accelerated implementation of moving least squares to interpolate space between the points. We also present evaluations of this approach both with a case study and with a user study.
}


\CCScatlist{ %
\CCScat{Computer Graphics}{I.3.6}{Methodology and Techniques}{Visualization}
}

\teaser{
\centering
\hbox{
\subfigure[PCA projection]{\includegraphics[width=5cm,height=5cm]{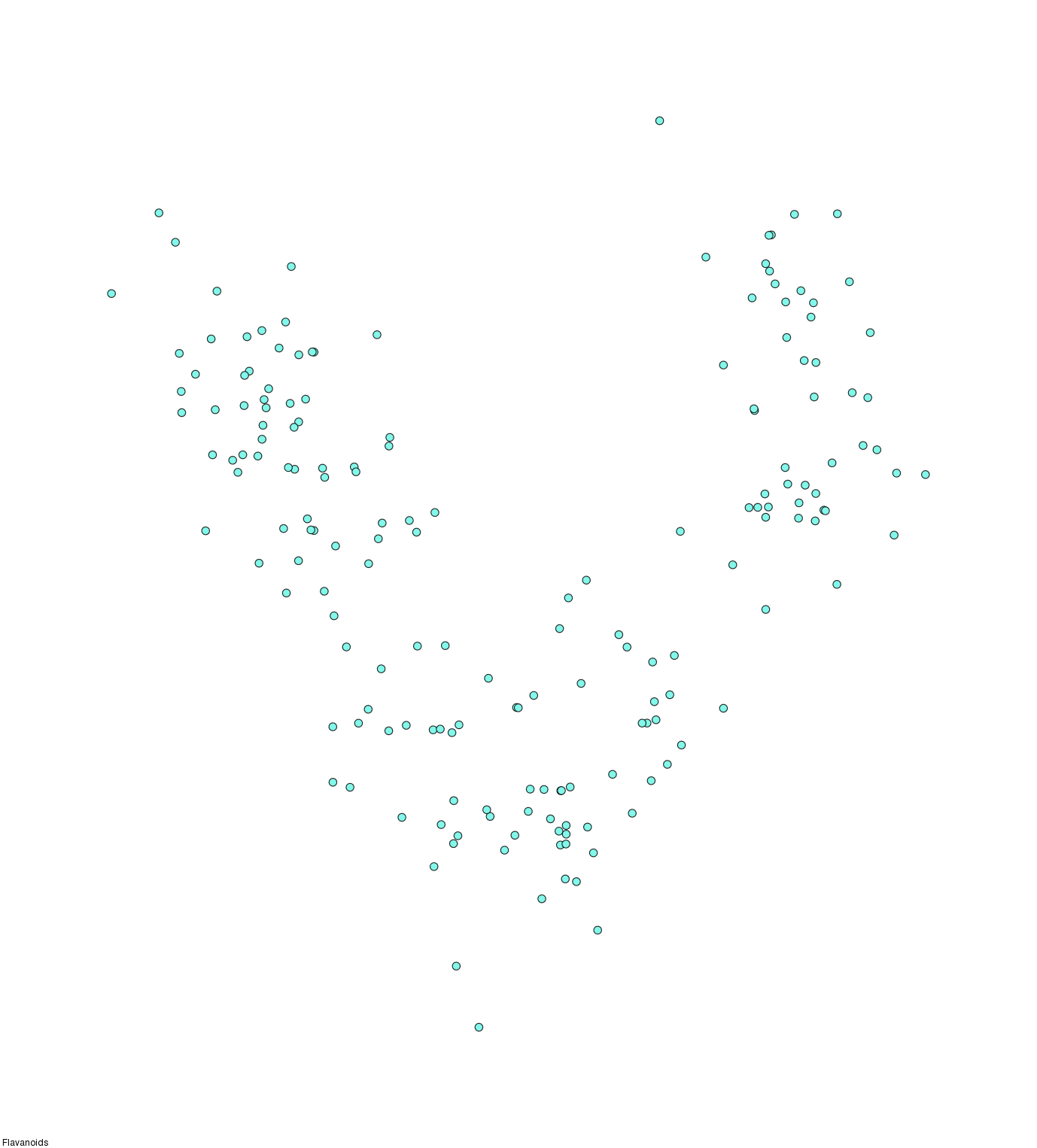}\label{teas1}}
\subfigure[Alcohol, discrete colors per contour]   {\includegraphics[width=5cm,height=5cm]{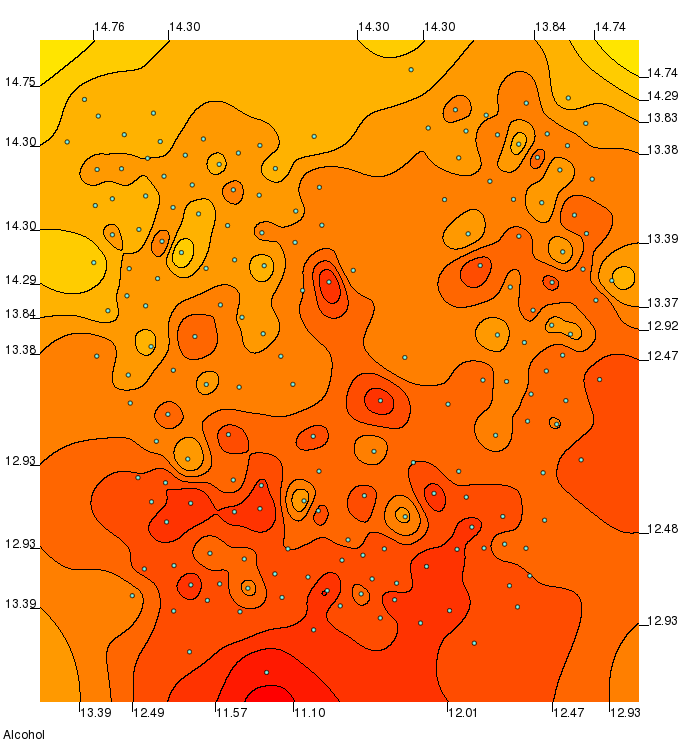}\label{teas2}}
\subfigure[Flavonoids, gradients between contours]   {\includegraphics[width=5cm,height=5cm]{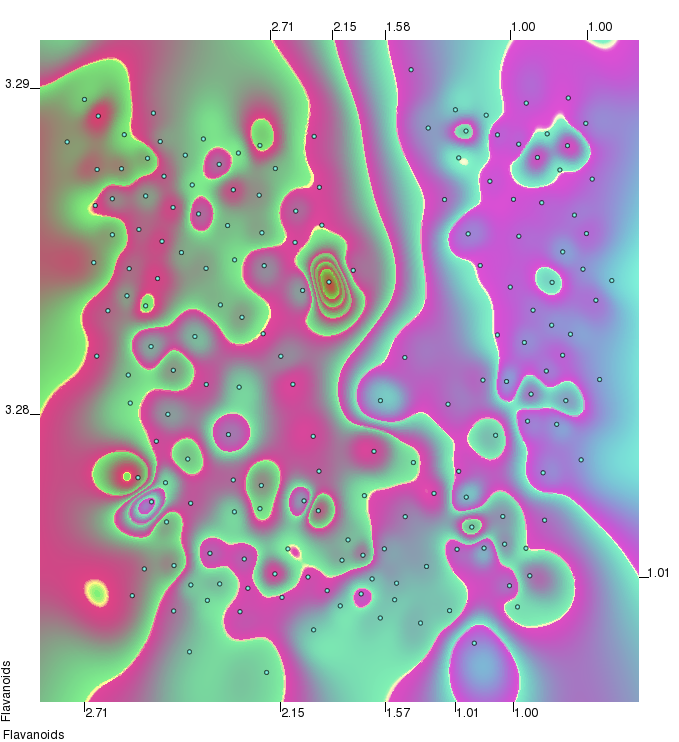}\label{teas3}}
}
\caption{
Traditional projection techniques can yield overplot and lose context of the original data dimensions (a).  By relaxing localized placement, dense points can be seperated,  while preserving neighborhoods.  Then, the background space can be used to convey a data dimension (b).  Gradients between the isocontours can provide further visual context of the precise value of points (c).
}
\label{teas}
}

\begin{document}

\firstsection{Introduction}

\maketitle

We live in a 3-dimensional world, so the human eye is quite attuned to perceiving and understanding at most 3 spatial dimensions at a time. But due to occlusion issues and since most modern displays are 2-dimensional, visualizations are frequently restricted to only 2 dimensions. As such, there are a multitude of techniques for visualizing data with 1 or 2 dimensions. Yet many data sets have more than 2 dimensions, and visualizing high-dimensional data is much more challenging. Non-spatial channels such as color or size can augment the points, but these often incur other perceptual issues. Most high-dimensional visualizations either display a few dimensions at a time or use some dimensionality reduction to project data down to 2 dimensions.

Visualizing 1 or 2 dimensions alone would not represent the entirety of the data, so many views of the same data are necessary.  This is commonly done with either small multiples or animated transitions. In the case of small multiples, many views are presented simultaneously. In order to correlate data points between the views, the views are often arranged such that they have 1 dimension in common. However, only that 1 dimension is shared, and additional dimensions can be difficult to track between plots, particularly if the data is dense. Also, if dimensions are compared pairwise, then the number of plots necessary increases quadratically with the number of dimensions (as in a scatterplot matrix), unless some of the combinations are skipped (as in parallel coordinates). This can quickly make such a visualization overwhelming, as well as require each plot to be quite compact. Animated transitions \cite{mdv-matdice} eliminate the compactness requirement by using the whole display to show selected dimensions at a time. However, animated transitions necessitate interaction, and so are not appropriate for all media. Also, unless paired with alternate views, they do not provide a sufficient overview.

An alternate approach to dealing with high-dimensional data is to reduce the dimensionality by projecting the data onto a lower dimensional space that can be represented with standard techniques (such as scatterplots). But simply projecting the data causes an inherent and unavoidable loss of information. There is also a risk of overplot, as many high-dimensional points can map to the same region of space (similar to occlusion in 3D applications), and many projections aim to project dense clusters in the data into small regions of space. This approach is also highly dependent on the choice of projection algorithm, as different projections will emphasize different aspects of the data. The resulting plot can also be unintuitive, as the projected dimensions can have little to no recognizable correlation with the original dimensions. Thus, even if a trend or pattern is clearly shown in a reduced dimensional plot, it can be difficult or even impossible to map that insight back to the input dimensions without additional information.  Simply coloring the points according to original dimension information can help with trend identification, but the human eye can not discern colors precisely enough for point valuation.  However, the human eye is much better at valuating spatial information.

In this paper, we describe a method of augmenting a reduced dimensional plot with isocontour-based representations to reincorporate the original dimensions. We begin with a Principal Component Analysis (PCA)-based dimensional reduction technique to create a 2D plot that retains direct correlations with the original dimensions. While this step presevers some aspect of the original dimensions, it can instantiate substantial overplot, so we then optionally apply a user-controllable amount of neighborhood-preserving distortion using a parallel constrained graph layout approach. And finally, we use a pixel-based Moving Least Squares (MLS) approach to compute texture coordinates for each pixel according to the original dimensions, and render the resulting field with one of several isocontour-based approaches, in order to utilize the eye's spatial valuation capabilities. The result is a plot with fixed points where the dimensions displayed are interchangeable, enabling direct comparison of different dimensions.

To summarize, the major contributions of this paper are:
\begin{itemize}
\item A parallel, planarity-preserving triangulation layout, including a mathematical proof of planarity and a GPU implementation.
\item A neighborhood-preserving, PCA-based dimensional reduction method which relaxes overly dense regions of the plot.
\item A per-pixel MLS based interpolation and rendering method for reintroducing the original dimensions.
\item An novel overall approach for multidimensional data visualization.
\end{itemize}

\section{Related Work} %
Multidimensional data is one of the fundamental classes of data \cite{datataxon}.  
As such, there is a large corpus of existing works on the visualization of multidimensional data \cite{mdv-stats,mdv-geom,mdv-taxonomy,mdv-tukey}.  These include glyph-based approaches \cite{mdv-cface,mdv-glyph,mdv-icon}, pixel based techniques \cite{mdv-pixel}, or stacked plots \cite{mdv-stack}, along with more geometric techniques such as scatterplots \cite{scatterplots, mdv-stats} and parallel coordinates \cite{mdv-parcoord}.  Scatterplot matrices \cite{mdv-matrix, mdv-matdice} are particularly common, as they show the relationships between all dimensions simultaneously.  However, they scale poorly with the number of dimensions as it takes $O(d^2)$ views to represent all the dimensions, so each individual view will be quite small.  Scagnostics can be used to single out plots of particular interest, but this would hide most of the dimensions, not to mention associations across plots would be even more difficult, nescessitating brushing and linking techniques.

Another common analysis method for multidimensional data is to project it down to a lower dimensional space that is easier to visualize \cite{pca,mds,lamp}.  Principal Component Analysis (PCA) \cite{pca} computes a linear system of orthogonal basis vectors to project the data onto in order to maximize the variance in each component.  Multi-Dimensional Scaling (MDS) \cite{mds} uses a matrix of computed similarities between items to compute points' locations.  Local Affine Multidimensional Projection (LAMP) \cite{lamp} lets the user place control points in space and projects the rest of the data around them.  In all these cases, the results are typically shown with simple visual representations, such as a scatterplot.

Self Organizing Maps (SOMs) have also been used to great effect for multidimensional data, where the points are organized in a space filling manner, and can then be colored according to the component planes \cite{som-org}.  However, as with simply coloring the points in a projection based method, the human eye is not precise enough to discern point values on color alone.

Many data projection and representation techniques can result in a high degree of overplot, where many data points map to the same area of the screen.
Data reduction techniques, such as sampling \cite{sampl1,sampl2} or clustering \cite{clust1,clust2}, can be used to address the issue of overplot by creating an abstracted overview, but in doing so, they do not show the entire dataset. In order to explore the entire dataset from such an overview, semantic zooming such as fisheye lensing \cite{fisheye0} with user interaction is necessary. Several works \cite{fisheye1,fisheye2,fisheye3,rubberplot} employ such fisheye lensing. Many of them do so purely geometrically without the need for semantic abstraction. However, they do not account for points that are precisely collocated. Keim et al. \cite{keim-gsp} address this issue by displacing overplotted points to the nearest free pixel. While this method does allow for all points to be drawn with minimum overplot, there is no visual indication of this displacement having occurred. Also, all of these methods use rectilinear distortion, which substantially limits their flexibility.

Nonrectilinear distortion has been used to great effect in other applications.  The work of Bak et al. \cite{density09} demonstrates a method of arbitrarily distorting maps so that geographic points are more evenly distributed.  Our approach can produce similar density equalizing distortion, but also aims to maintain the capability to quantitatively valuate point locations by smoothly interpolating the surrounding space. 

One step of our approach involves computing a new layout for a triangulated graph. There are many existing algorithms for the layout of general graphs \cite{graphsurvey}. 
Our case is slightly specialized in that we start with a planar graph, and guarantee planarity at each step of the layout algorithm. The approach of Dwyer et al \cite{topograph} preserves planarity, but does so by allowing for non-linear edges. Some layouts are force-directed approaches that preserve edge-crossing properties, but which make assumptions that are only true in a serial implementation \cite{bertault_force-directed_1999,impred}. 
The force-directed algorithm we use is a modified GPU version of
FM$^{3}$ \cite{hachul_drawing_2005}, which is similar to other GPU
implementations \cite{frishman_multi-level_2007,godiyal_rapid_2009}. The primary difference between our layout and these existing implementations of FM$^{3}$ is that our layout imposes a planarity-preserving constraint.

There are many methods for interpolating between unstructured data points, including Barycentric/linear interpolation, Sibson interpolation \cite{sibson}, and Moving Least Squares (MLS) \cite{mls-asap}.  Of these, MLS offers the smoothest continuity.  While MLS is often used in 3D applications such as surface reconstruction \cite{mls-3drob}, it has also been applied to 2D applications such as image deformation, as in the work of Schaefer et al. \cite{imagemls} where MLS is used to distort a grid of points according to a skeleton of control points. MLS has also be used to interpolate internal coordinates for higher order polygons \cite{mls-coords}. Our approach uses the deformation calculations from Schaefer et al. \cite{imagemls}, but uses them to interpolate texture coordinates as in \cite{mls-coords}.  Also, we use GPU acceleration to parallely perform the MLS calculation per pixel.  

Finally, the space rendering techniques we employ produce isocontour maps. This is a common representation also employed by topographic maps \cite{topomap}.  Isocontours are well understood to be legible and clear, even for multiple dimensions, and are familiar to most users.

\section{Methodology} %
The primary goal of this work is to 
present a point cloud using the background space to convey the quantitative values.
Doing this mandates that we maximize the available background space around each point (or at least around each cluster of similar points).
Relatedly, we also want to preserve locality, so that neighboring points are also neighboring in the original space.
At the same time, we want to preserve readability and be able to convey the same information as a traditional plot, such as clusters and large scale trends.

\begin{figure}
\centering

\subfigure[Initial projection]{\includegraphics[width=.3\linewidth, width=.3\linewidth]{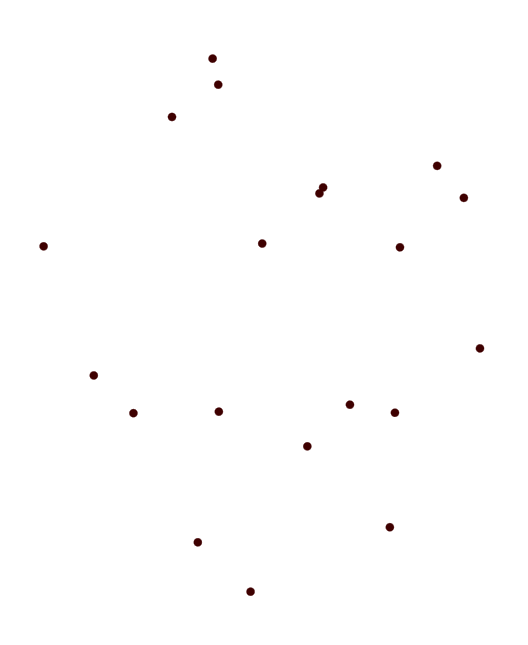}\label{fig-steps-init}}
\subfigure[Direct MLS]{\includegraphics[width=.3\linewidth, width=.3\linewidth]{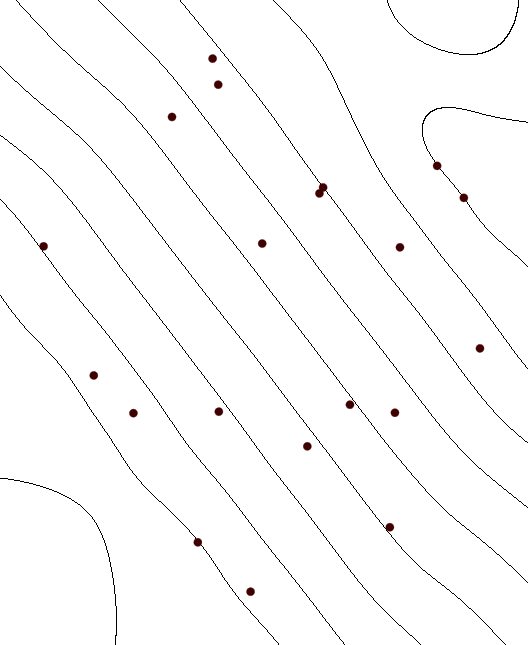}\label{fig-steps-mlsa}}
\subfigure[Triangulate]{\includegraphics[width=.3\linewidth, width=.3\linewidth]{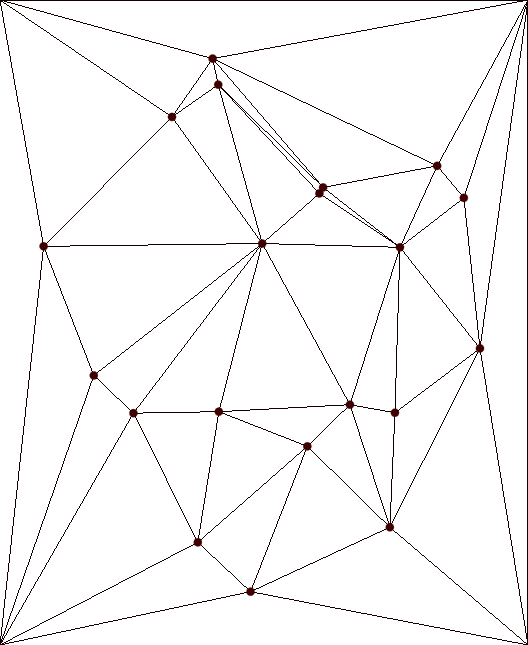}\label{fig-steps-triangle}}
\subfigure[Relayout]{\includegraphics[width=.3\linewidth, width=.3\linewidth]{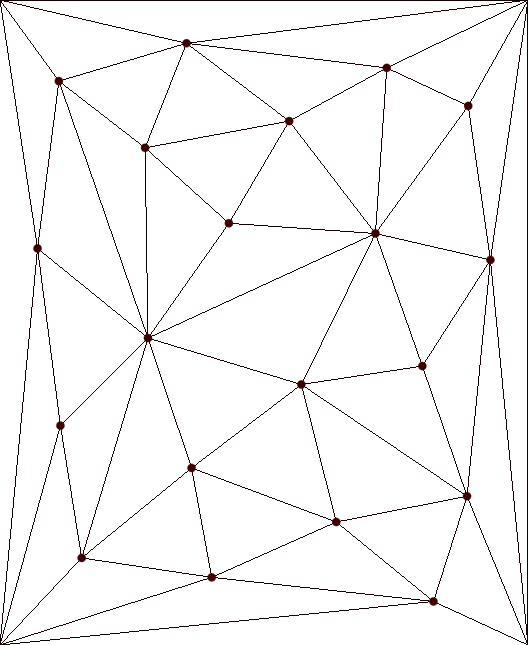}\label{fig-steps-fdl}}
\subfigure[MLS contours]{\includegraphics[width=.3\linewidth, width=.3\linewidth]{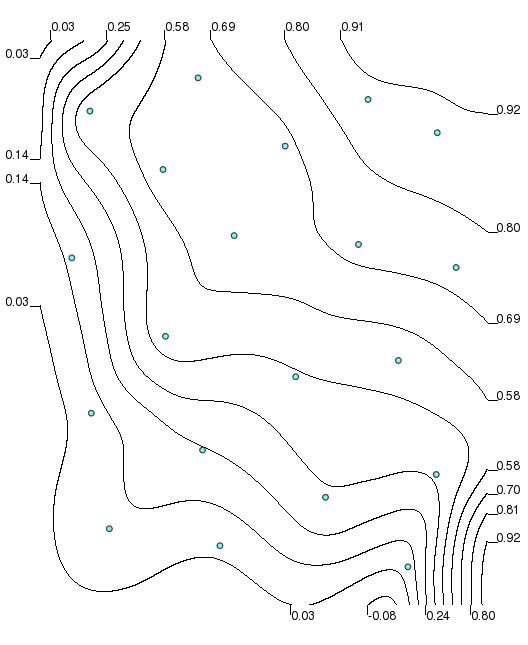}\label{fig-steps-mlsd1}}
\subfigure[MLS gradients]{\includegraphics[width=.3\linewidth, width=.3\linewidth]{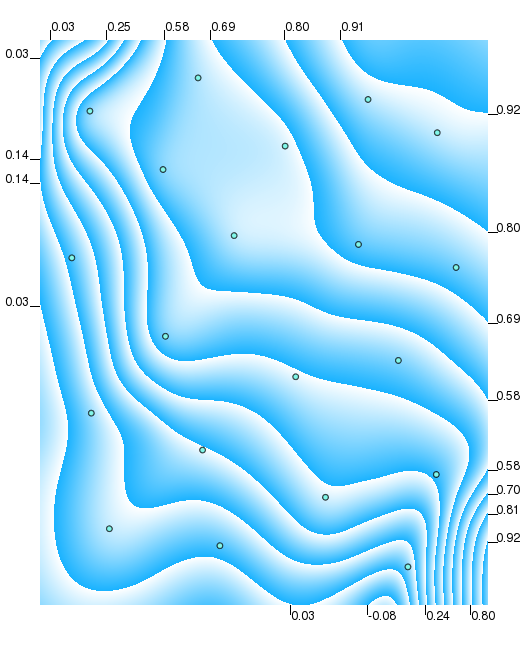}\label{fig-steps-mlsd2}}

\caption{
\emph{Overall process.} From the initial projection (a), we can apply MLS interpolation directly (b), or we can triangulate the points (c), shift them towards a constrained force-directed layout (d), and then use MLS to interpolate and render input dimensions in the space between the distorted points (e,f).
}
\label{fig-steps}
\end{figure}

Our overall process is depicted in Figure~\ref{fig-steps}.
Unless the data is already 2-dimensional, we first project the data down to 2 dimensions, as shown in Figure~\ref{fig-steps-init}.  Then, from the resulting plot, we capture the initial layout and neighborhood information by computing
a Delaunay triangulation of the projected points (Figure~\ref{fig-steps-triangle}). We treat the triangulation as a graph, and compute a target layout using a parallel, planarity-preserving, force-directed layout algorithm (Figure~\ref{fig-steps-fdl}).
The user is allowed to manually interpolate between the original layout and the relaxed layout to find a preferable arrangement of the points.
Finally, we use MLS to interpolate space between the points, and render one or more of the original dimensions with one of several possible mappings (Figure~\ref{fig-steps-mlsd1} and \ref{fig-steps-mlsd2}).

\subsection{Triangulation} %
Many dimension-reduction projections (including PCA) often place many points very close together, or even on top of each other.  This can make it difficult to evaluate the membership of clusters, and can leave little to no room for our approach to render the contextual space.
Rectilinear distortion \cite{keim-gsp, fisheye1} can spread points out, but cannot use space efficiently, as it rigorously preserves ordering in both the X and Y dimensions.
But since the dimensions in a multidimensional plot are not directly meaningful, there is little reason to preserve rigorous ordering.
What is meaningful in such a projection is the local relationships between points.
So we want to relax the point locations,  but preserve localized neighborhoods (i.e. points that are neighbors in the original space will still be neighbors after distortion). One way of defining these neighbors is by triangulating the points to create a mesh of neighborhood connections.
While any triangulation would work, we choose to use a Delaunay triangulation, as its relation to Voronoi diagrams means that it carries a connotation of nearest neighbor calculations. It also has several nice geometric properties. Namely, the expected number of triangles and edges around any given node is constant, and the avoidance of skinny triangles means that average initial edge length will be minimized.  The amortized number of triangles around each point is useful for guaranteeing complexity of subsequent calculations. Additionally, the minimal average edge length will reduce the distance nodes have to move to optimize the mesh layout.

\subsection{Planarity Preserving Triangular Mesh Layout} %
Most graph layout algorithms start from the assumption that there is no initial positioning for the nodes \cite{graphsurvey}. As such, they have to deal with untangling randomly placed nodes, but also have the freedom to move nodes anywhere. In our case, the initial layout given by the plotted points is planar, so our layout method does not need to spend any iterations on untangling the network. However, node motion needs to be limited in order to maintain planarity at all times. \cite{bertault_force-directed_1999} and \cite{impred} describe methods to guarantee planarity by constraining each node's motion in a force directed layout such that no individual move violates the planarity. While these approaches are sufficient in a serial implementation, they will not work in a parallel implementation, such as one performed on the GPU.  When multiple nodes' motions are calculated in parallel, they cannot incorporate their neighbors' changes, so it is possible for several independently allowable moves to violate planarity when performed in combination. Here we describe a new constrained graph layout method for triangular meshes that can be performed in parallel on the GPU while still guaranteeing planarity.

\subsubsection{Force-Directed Layout} %
Our layout is based on a GPU implementation of FM$^3$, similar to \cite{frishman_multi-level_2007,godiyal_rapid_2009}.
We eliminated the multipole portion of the algorithm for the sake of the planarity guarantee. 
This slows the convergence of the graph, but the overall impact is minimal, since the graphs we handle are fairly small, and start planar already. In practice, any plot with enough points to significantly slow down the layout would likely be too dense to display on the screen all at once, even with our method.  

The initial force model acting on each vertex $v$ consists of a sum of repulsive inverse square forces $\vec{F_r}$ from every other vertex and non-physical spring forces $\vec{F_e}$ from $v$'s neighbors, defined as: 

\begin{equation}
  \vec F_r( v,  v_i |  v_i \ne  v) = \frac{ C } { {\| \vec V \|}^3 + \eta } \vec{V},~
  \vec V =  v -  v_i
\end{equation}

\begin{equation}
 \vec F_e( v,  v_i | [  v,  v_i] \in E) = {\| \vec V \|}
  \log{ \frac{\| \vec V \| + \eta }{D} } ( -\vec V ), ~
  \vec V =  v -  v_i
\end{equation}

\noindent
where the constants $C$ and $D$ correspond to the repulsive scaling constant and the desired edge length, respectively. $\eta$ is used throughout as a softening factor to avoid numerical error. We add to these forces a node-edge repulsive force that operates between a node and its opposing edge for each neighboring triangle. This force is calculated as:

\begin{equation}
 \vec F_{re}( v_i,  v_j | [  v,  v_i,  v_j] \in T) = \frac{ -C }
  { \| \vec R \|^2 + \eta } \vec {R}
\end{equation}

\noindent
where $\vec{R}$ is the vector from the point $v$ to the line $ v_iv_j$. Simulated annealing is used for graph convergence. The movement of each node during an iteration $n$ is limited by the temperature $t_n$. Given a tuneable initial temperature $t_i$, the temperature decays according to the decay constant $\lambda$, such that $t_n = {\lambda}^n t_i$.

The original force model acts to distribute nodes evenly and make the average edge length $D$. In theory, this should create regular triangles, but in practice, we have found that many long, narrow triangles persisted. Triangles with such sharp angles create high-frequency distortion and make the contour plot more difficult to read. The additional node-edge repulsive force acts to make triangles more regular and reduces the occurrence of sharp angles. 

The GPU implementation uses a kd-tree for approximating the repulsive force of distant groups of nodes as point charges. A Compressed Sparse Row (CSR) representation is used to store edges because of its storage efficiency and contiguous memory allocation. Our CSR consists of two one-dimensional arrays - one to store an index for each node, indicating the start of its edge list, and another to store a destination index for each edge. 

A storage scheme similar to CSR is used to store the triangulation as triangle fans. Storing triangles in this manner only requires $|N| + |T|$ extra space. Since the Delaunay triangulation yields a constant number of triangles on average per vertex, the triangle data only adds a small linear scaling factor to overall space required by the layout algorithm. The guarantee on the average number of triangles means that the node-edge force only adds a small linear factor to the running time of the layout algorithm. Thus, no special optimization is necessary to compute the node-edge repulsive force.

\begin{figure}[t]
\centering
\subfigure[Initial triangle and limiting areas]{\includegraphics[width=.49\linewidth, keepaspectratio]{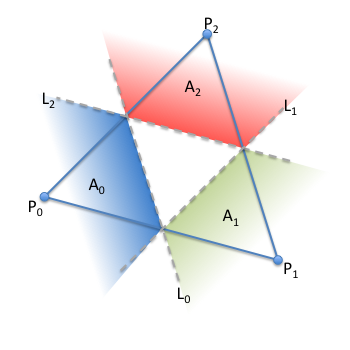}\label{fig-proof-init}}
\subfigure[Moved points with vectors, angles, and intersections]{\includegraphics[width=.49\linewidth, keepaspectratio]{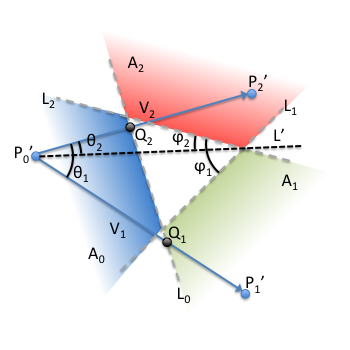}\label{fig-proof-end}}
\caption{\emph{Diagrams for planarity proof.} After the move, the angles between any two sides are kept within the range $(0,\pi)$, so the sign of the signed area is preserved. Therefore, the triangle cannot flip.}
\label{fig-proof}
\end{figure}

\subsubsection{Planarity Constraint} %
In order to guarantee that planarity is preserved throughout the layout process, it is sufficient to show that planarity is preserved for each iteration. In order for planarity to be broken, some node must cross over some edge. If this occurs, then the triangle formed by that node and edge would flip, and the sign of its signed area would change. Thus, if we can prevent any triangles from flipping, we can guarantee that the layout will remain planar.

At each iteration of our algorithm and for each triangle, we construct three limiting lines running through the midpoints of each edge triangle, as shown in Figure \ref{fig-proof-init}. In a single iteration, each vertex is constrained to never cross any of the limiting lines of any of its triangles. We will show that this is sufficient to prevent the triangle from changing orientation during an iteration even when all 3 nodes move simultaneously, and therefore guarantee that the triangle has the same orientation at the end of the algorithm as at the start.

Given any triangle with non-zero area, arbitrarily choose a point $P_0$ and assign the points $P_1$ and $P_2$ in counter-clockwise order so that the signed area will be positive (i.e. the angle from $P_1 - P_0$ to $ P_2 -  P_0$ is in the range $(0,\pi)$). Then the limiting lines $L_0$, $L_1$, and $L_2$ and the areas of allowed movement $A_0$, $A_1$, and $A_2$ for each point are defined as in Figure~\ref{fig-proof-init}. Now choose any $ P_0' \in A_0$, $ P_1' \in A_1$, and $ P_2' \in A_2$, and let $\vec V_1 =  P_1'- P_0'$ and $\vec V_2 =  P_2'- P_0'$ The signed area of the new triangle $ P_0' P_1' P_2'$ can then be calculated as $SA=|\vec V_1||\vec V_2| \sin \theta$, where $\theta$ is the angle from $\vec V_1$ and $\vec V_2$. The sign of this area is positive if and only if $0<\theta<\pi$ 

Let $L'$ be the line that goes through $ P_0'$ and the intersection of the limiting lines $L_1$ and $L_2$, as shown in Figure~\ref{fig-proof-end}. Since $ P_0'$ cannot cross $L_1$, the angle $\phi_1$ between $L'$ and $L_1$ must be greater than $0$. Likewise, $\phi_2$ between $L'$ and $L_2$ must be greater than $0$. As such, $L'$ can intersect neither $A_1$ nor $A_2$. Therefore, the angle $\theta_1$ between $\vec V_1$ and $L'$ must be greater than $0$ regardless of the selection of $ P_0'$ or $ P_1'$. The angle $\theta_2$ must also be greater than $0$. Thus, $\theta=\theta_1+\theta_2>0$. 

Now consider the intersections of the vectors $\vec V_1$ and $\vec V_2$ with the line $L_0$. Let the points $Q_1$ and $Q_2$ be these intersections, as shown in Figure~\ref{fig-proof-end}. Since $\theta>0$, $Q_1 \ne Q_2$. If $\theta=\pi$, then the points $Q_1$, $Q_2$, and $P_0'$ would be collinear. As $Q_1$ and $Q_2$ both lie on the line $L_0$, then $P_0'$ would be on the line $L_0$ as well. However, $P_0'$ cannot be on the line $L_0$ by definition. Therefore $\theta$ cannot be $\pi$, so it is constrained to the range $(0,\pi)$, meaning that the signed area of the new triangle is positive and thus the triangle did not flip, which proves that planarity is preserved.

In practice, the planarity constraint looks similar to the node-edge force. For a single $\vec v$, we calculate the vectors $\vec R_i$ from $\vec v$ perpendicular to each of the limiting lines $L_0$, $L_1$, and $L_2$. $\vec R_i$ is used to calculate the vector projection of the force $\vec F_R$, and limit it such that $\| \vec F_R \| \leq \| \vec R_i \|$ for $i = [0, 1, 2]$. This adds another linear scaling factor to the overall running time, but again, this is a small factor due to the guarantees of the initial Delaunay triangulation.

The constraints break down if the vertices become co-linear and the triangle area becomes zero. In theory, a triangle can only asymptotically approach zero area, but after a large number of iterations numerical error can push a node past an edge. We use the softening factor $\eta$ to limit vertices from coming within $\eta$ of any limiting line.

\begin{figure}[t]
\centering
\subfigure[Linear Interpolation]{
\includegraphics[width=.48\linewidth, width=.48\linewidth]{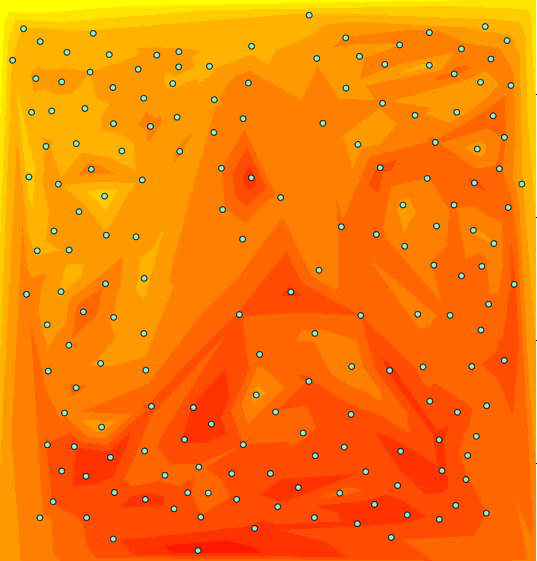}\label{fig-mls-lerp}}
\subfigure[Mean MLS]{
\includegraphics[width=.48\linewidth, width=.48\linewidth]{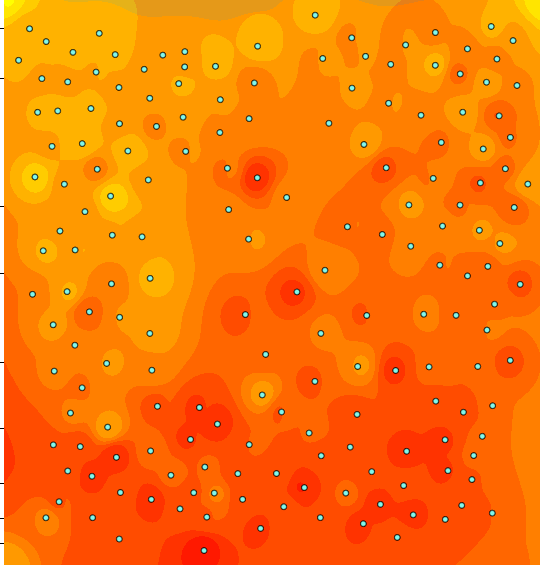}\label{fig-mls-mean}}
\subfigure[Affine MLS]{
\includegraphics[width=.48\linewidth, width=.48\linewidth]{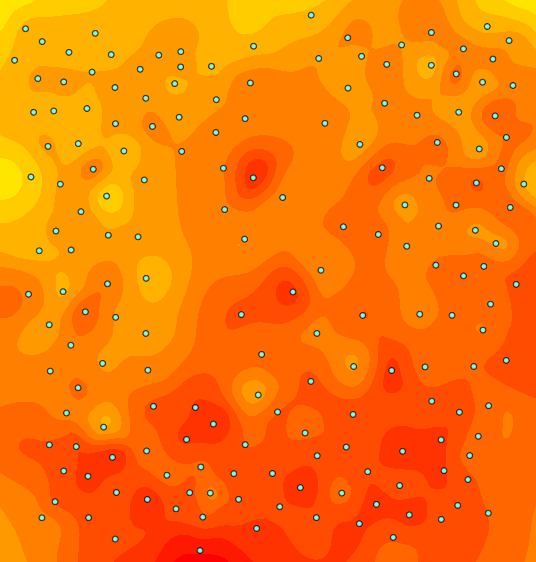}\label{fig-mls-affine}}
\subfigure[Rigid MLS]{
\includegraphics[width=.48\linewidth, width=.48\linewidth]{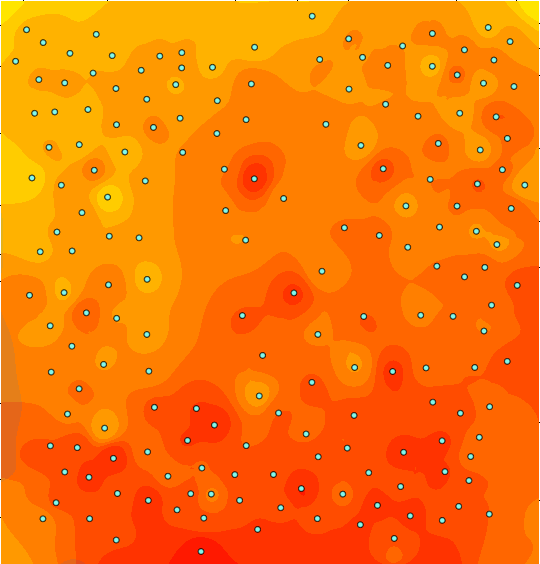}\label{fig-mls-rigid}}
\caption{\emph{Moving Least Squares variants.} Standard linear interpolation (a) is insufficient, as the sharp boundaries are very noticeable. Mean MLS (b) is the most efficient MLS, but is less rigorously defined. Affine MLS (c) allows only affine transformation, but can still allow for non-uniform scaling and shearing. Rigid MLS (d) improves upon this, and ends up being very similar to Mean MLS.}
\label{fig-mlsvar}
\end{figure}

\subsection{Moving Least Squares} %
Once the points have been distorted, the space between them must be evaluated. As the space between the points is already broken down into triangles, the most straightforward way to interpolate space continuously is to render each triangle individually and simply linearly interpolate between the three points with barycentric coordinates, as in the standard graphics rendering pipeline. But linear interpolation only provides $C^0$ continuity, leading to sharp angles and obvious triangle boundary artifacts which detract from readability, as can be seen in Figure~\ref{fig-mls-lerp}.   Splining techniques could help resolve these sharp angles, but would only be easily applied to contour lines, not more complex rendering methods that require interpolation between contour lines.  Also, the splining techniques would be restricted to ones that can handle closed loops, and which can be restricted so they do not cross over other lines or points.  

For smooth distortion, we require higher order interpolation involving more than just the three points of each triangle. Specifically, we need a function $\vec f(\vec v)$ to map points $\vec v$ from the space containing the distorted points $\vec p_i$ to the space containing the original points $\vec q_i$ and which meets the following criteria:
\begin{enumerate}
    \item Smoothness: $\vec f(\vec v)$ is $C^1$ continuous or better
    \item Point identity: $ \vec f(\vec p_i ) = \vec q_i$
    \item Smooth even with no distortion:
        $\forall i (\vec p_i = \vec q_i) \Rightarrow \vec f(\vec v)=\vec v$
\end{enumerate}
Moving Least Squares (MLS) meets these criteria \cite{imagemls}. 

MLS is a radial basis function based mapping, where the contribution applied to a given sample from the distortion of each control point is given weight inversely proportional to the distance from the sample to the control point. As in \cite{imagemls}, we use weights $w_i$ of the form
\begin{equation}
w_{i}(\vec v)=\frac{1}{|{\vec p_{i}-\vec v}|^{2\alpha}} 
\label{eq-weights}
\end{equation}
It can be easily seen that as $\vec v \rightarrow \vec p_i$, $w_i \rightarrow \infty$.   This property guarantees that distorted points map to their original positions. We let $\alpha$ be a user-tunable constant, but find the range $.5<\alpha<1.5$ to work well, depending on the MLS mode. 

Rather than distorting points from the original space into the distorted space as in \cite{imagemls}, we map each pixel in the distorted space to their texture coordinates in the undistorted space. By applying MLS this way, it is natural to use a fragment shader to accelerate the MLS calculation with the GPU. We do this in 2 passes. The first pass performs MLS to map each pixel to a location in the original space and renders the coordinates to a frame buffer object.  Then, the second pass takes the distorted coordinate texture and applies a visual mapping to present the distorted space to the user.

MLS can use many different functions weighted on the distance to the control points. We use a simple function we call Mean MLS (Figure~\ref{fig-mls-mean}), and two of the three transformation functions from \cite{imagemls}: Affine MLS and Rigid MLS (Figures~\ref{fig-mls-affine} and \ref{fig-mls-rigid} respectively). The third function from \cite{imagemls}, Similarity MLS, was not implemented due to its partial terms which would require a third pass to perform efficiently on the GPU.

\subsubsection{Mean Moving Least Squares} %
Unlike many applications of MLS, we do not nescessarily require the transformation to be affine. As such, the simplest method of MLS we implemented calculates the weighted average of the displacement from the distorted points' positions to their original positions. That is, we calculate
\begin{equation}
\vec f_m(\vec v)=\vec v + \frac{\sum_{i}(w_i(\vec v)*(\vec q_i-\vec p_i))}{\sum_{i}w_i(\vec v)} 
\label{eq-meanmls}
\end{equation}
where $w_i$ are the weights defined in Equation \ref{eq-weights}, $\vec p_i$ are the points' distorted positions, and $\vec q_i$ are the points' original positions. Given that the weights are inversely proportional to the distance to each distorted point, then as $\vec v \rightarrow \vec p_i$, $w_i \rightarrow \infty$,  so $\vec f_m(\vec v) \rightarrow \vec v + \vec q_i-\vec p_i = \vec q_i$. Also, with the exception of the exact locations of the points, $\vec f_m(\vec v)$ is smooth and continuous. This choice of function also preserves undistorted space when the control points are not moved, as $\forall i (\vec p_i = \vec q_i)  \Rightarrow \vec f_m(\vec v) = \vec v + 0 = \vec v$. Thus, even though the calculation is simple, it meets our legibility criteria. Figure~\ref{fig-mls-mean} demonstrates this distortion, showing that it is much smoother and more legible than standard linear interpolation.

\subsubsection{Affine Moving Least Squares} %
While Mean MLS naively addresses the readability requirements and is computationally efficient, it was not derived rigorously and thus we have not proven that the mapping covers the entire original space, nor that the distortion is optimal. 
\cite{imagemls} defines three affine MLS transformations that make these guarantees, along with the derivations that prove their optimality. Of these three, we have implemented two. The first is a general affine MLS, which is calculated as follows:
\begin{equation}
\vec f_a(\vec v)=(\vec v - \vec p_*)\left(\sum_i (w_i \vec p_i^T \vec p_i)\right)^{-1} \sum_i (w_i \vec p_i^T \vec q_i) + \vec q_*
\label{eq-affinemls}
\end{equation}
where $\vec p_*$ and $\vec q_*$ are the weighted centroids defined as
\begin{equation}
\vec p_*=\frac{\sum_{i}(w_i(\vec v)*(\vec p_i))}{\sum_{i}w_i(\vec v)} 
\label{eq-pstar}
\end{equation}
\begin{equation}
\vec q_*=\frac{\sum_{i}(w_i(\vec v)*(\vec q_i))}{\sum_{i}w_i(\vec v)} 
\label{eq-qstar}
\end{equation}

As described in \cite{imagemls}, $\vec f_a(\vec v)$ optimizes the mean square error while restricting the deformation to affine transformations only.  This guarantees that it will cover the entire space and that it is optimal. 

\subsubsection{Rigid Moving Least Squares} %
While the affine transformation is mathematically smooth, it can lead to shearing and non-uniform scaling, which, \cite{imagemls} argues, can be less visually appealing. The rigid MLS transformation from \cite{imagemls} addresses this by further limiting the transformation to only translation and rotation. The rigid MLS transformation is calculated as
\begin{equation}
\vec f_r(\vec v)=| \vec v - \vec p_* | \frac {\vec f_r'(\vec v)}{| \vec f_r'(\vec v) |} + \vec q_*
\label{eq-rigidmls}
\end{equation}
where $\vec f_r'( \vec v )$ is defined as:
\begin{equation}
\vec f_r'(\vec v)= \sum_i w_i \Delta \vec q_i 
\left(
\begin{array}{c}
\Delta \vec p_i  \\
-\Delta \vec p_i^\bot
\end{array}
\right) 
\left(
\begin{array}{c}
\Delta \vec v_i \\
 -\Delta \vec v_i ^\bot
\end{array}
\right)^T 
\label{eq-rigidmlspartial}
\end{equation}
with $\Delta \vec q_i = \vec q_i - \vec q_* $, $\Delta \vec p_i =  \vec p_i-\vec p_*$, and $\Delta \vec v_i =  \vec v_i-\vec p_*$, and $w_i$, $\vec p_*$, $\vec q_*$ defined as in equations \ref{eq-weights}, \ref{eq-pstar}, and \ref{eq-qstar} respectively. 

The full derivation of this transformation can be found in \cite{imagemls}. Because it involves the calculation of a square root, it is more computationally expensive than the other methods. But in our experience, the resulting deformation is quite close to the results from mean MLS, as can be seen in Figures \ref{fig-mls-rigid} and \ref{fig-mls-mean}.

\begin{figure}[t]
\centering
\subfigure[Contour Lines]
{\includegraphics[width=.48\linewidth, height=.48\linewidth]{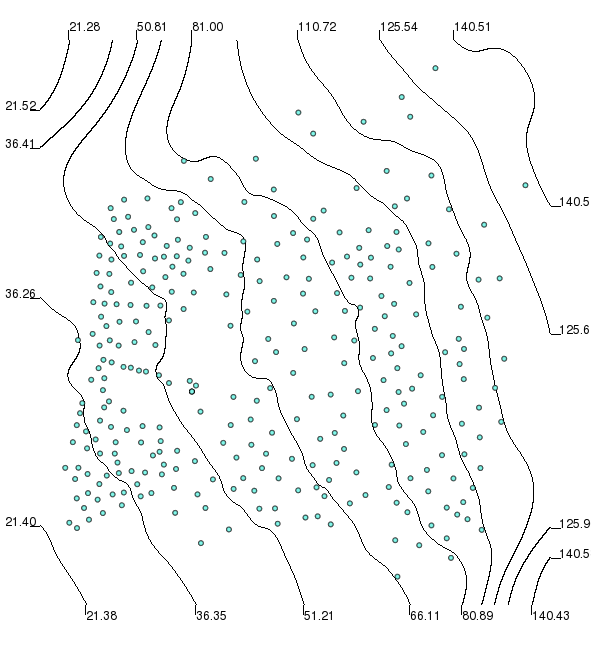}\label{fig-render-grid}}
\subfigure[Discrete Colors]
{\includegraphics[width=.48\linewidth, height=.48\linewidth]{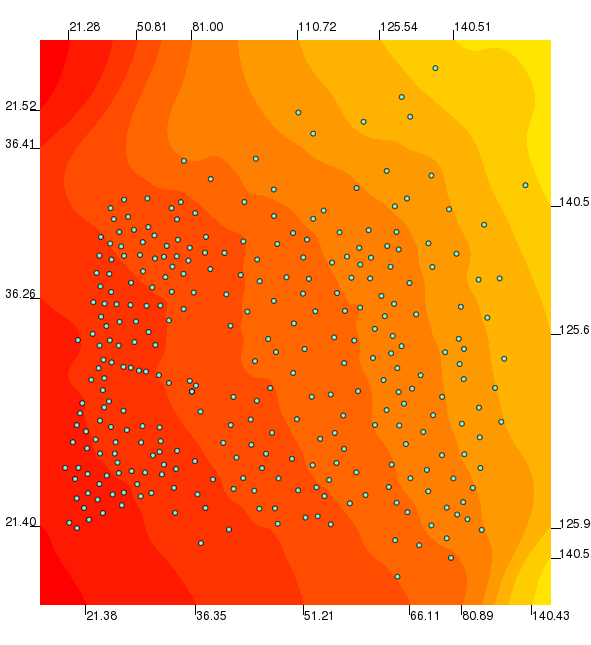}\label{fig-render-discrete}}
\subfigure[Discrete Colors + Contour Lines]
{\includegraphics[width=.48\linewidth, height=.48\linewidth]{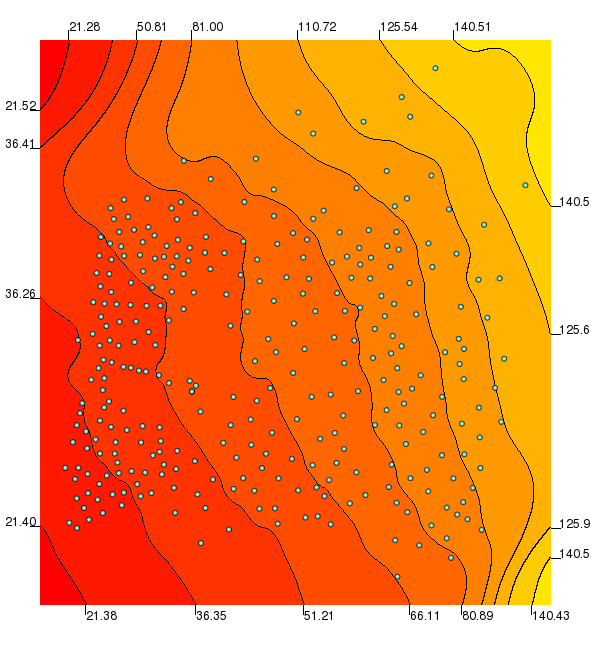}\label{fig-render-griddisc}}
\subfigure[Adaptive Contour Lines]
{\includegraphics[width=.48\linewidth, height=.48\linewidth]{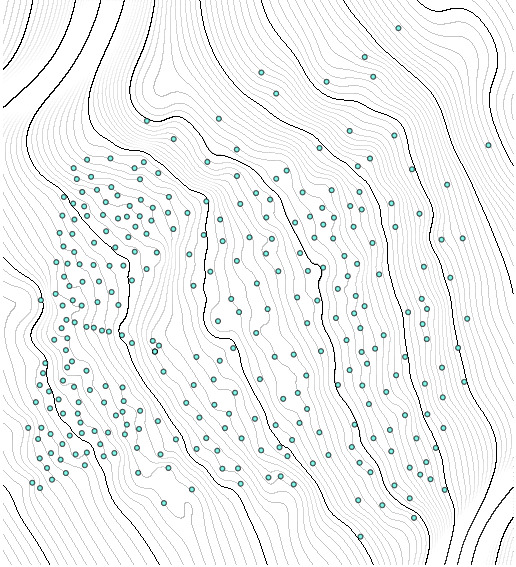}\label{fig-render-adaptive}}
\subfigure[Gradients]
{\includegraphics[width=.48\linewidth, height=.48\linewidth]{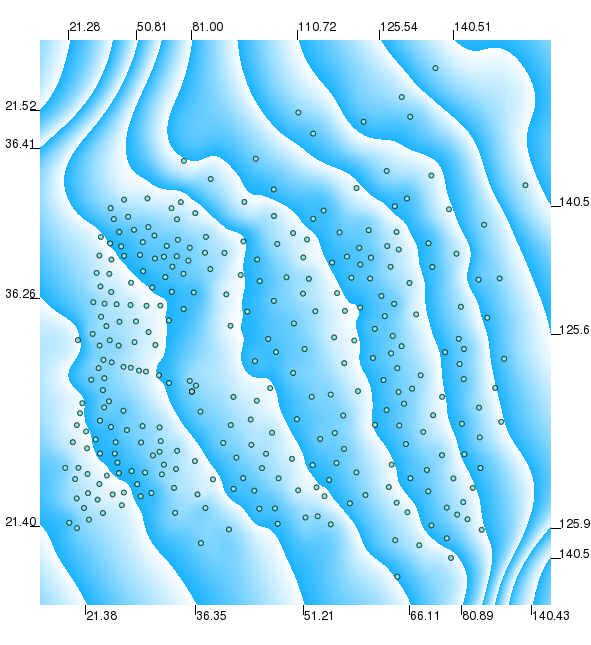}\label{fig-render-grad}}
\subfigure[Texture Map]
{\includegraphics[width=.48\linewidth, height=.48\linewidth]{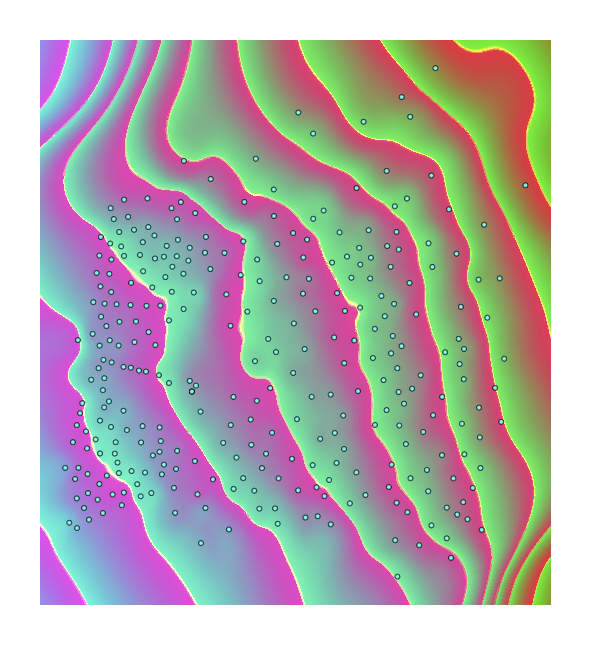}\label{fig-render-texture}}
\caption{\emph{Space Mapping Methods.} The distorted texture coordinates are rendered in one of several human interpretable ways.}
\label{fig-render}
\end{figure}

\subsection{Space Mappings} %
Once the texture coordinates have been calculated by MLS, they need to be presented visually to the user. There are many different ways in which this can be done. Just as the simplest method for showing a background for a scatterplot is to provide gridlines, the simplest method for rendering the MLS coordinates is to draw isocontours, as in Figure \ref{fig-render-grid}. This representation is very similar to the one used by topographic maps to display different elevation levels. While simple and traditional, there is no inherent ordering to the lines, meaning that it is not possible to tell whether a closed loop corresponds to a local minimum or maximum.

One way to address this limitation is to use color. By assigning a discrete color to each region of the plot, as in Figure \ref{fig-render-discrete}, the ordering is given by the colormap, and the boundaries can be seen as differences in color. However, as the number of regions increases, color differences can become too small to distinquish. If we combine the discrete colors and the contour lines we can address both of these issues simultaneously, as shown in Figure \ref{fig-render-griddisc}. 

However, in all three of these representations it can be difficult to concretely evaluate a point's actual value, as space is non-linear. One way to address the lack of information is to add additional lines adaptively, as in Figure \ref{fig-render-adaptive}.  In this method, we evaluate the gradient of the distorted space to determine the localized spatial resolution, then fade in more lines in areas of the plot where space is expanded and fade out excess lines where space is compressed. Another way to address this is to add gradients between contours, as in Figure~\ref{fig-render-grad}. In this manner, the color more precisely defines the value, and the gradient direction defines an ordering (in this example, blue to pink is increasing x, and blue to cyan is increasing y).  Finally, we also include the option to use the distorted coordinates to map an arbitrary texture to space, as shown in Figure \ref{fig-render-texture}, which allows for additional data-specific cases that may arise, or any user-definable representations that could help, such as a texture that defines both overall ordering as well as localized gradients.

\subsection{Interaction}
One advantage to our approach is that it requires very little interaction to generate the final plots. Specifically, the user only needs to choose a variant of MLS and rendering mode, and then adjust the MLS exponent ($\alpha$). The first two options are handled in our implementation via radio-boxes, and the latter is adjusted by a slider. Selecting a good exponent is important, as too small of an exponent will not interpolate enough between points, and too large of an exponent can introduce non-aesthetic artifacts. Each of the MLS variants behaves slightly differently, so the ideal exponent is slightly different for each, with the mean and rigid MLS variants generally requiring a smaller exponent than the affine MLS variant.
Transitioning back and forth from the original projection to the distorted projection is also done with a slider, allowing the user to scrub back and forth between the two, and control how much they distort the layout. 
Because most of the computation is on the GPU, the rendering is fast enough that these parameters can be adjusted interactively. 
The user generally does not need to tune the parameters much, as we have preselected good defaults.  Finally, we expose the parameters of the graph layout algorithm to the user, allowing the user to tune and recompute the layout if needed. However, the graph layout is relatively computationally expensive, so it is only run on demand.

Once the points are laid out and the MLS is tuned, the user can switch between showing one or two of the data dimensions at a time, and select which dimension(s) to use.  The user can quickly and easily cycle through the dimensions.  Since the points do not move, context is preserved between plots without the need for animated transitions.

\begin{figure*}
\subfigure[Scatterplot Matrix]{{\includegraphics[width=.6\linewidth, height=.6\linewidth]
{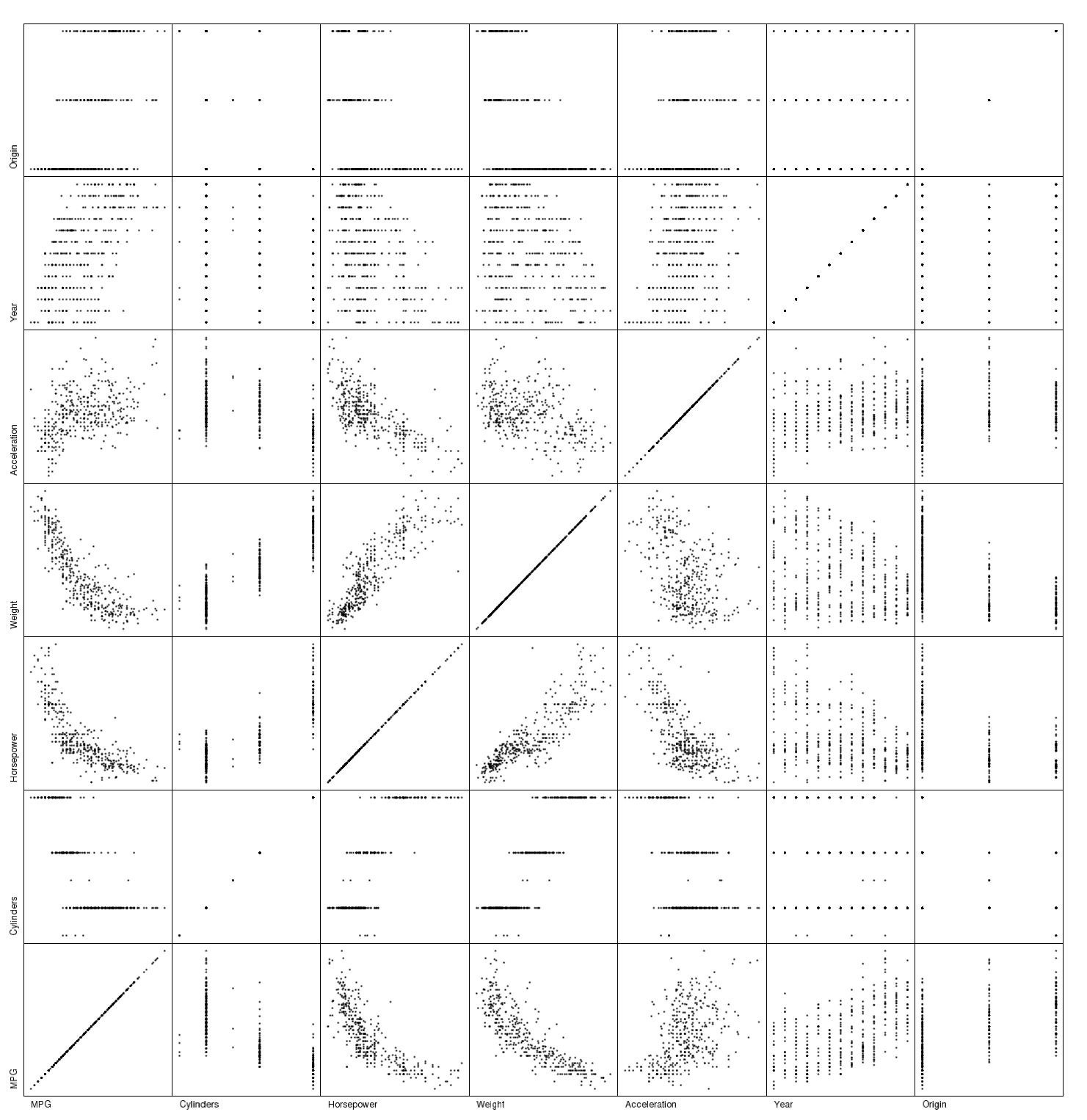}}        \label{fig-car-mat}}
\subfigure[Parallel Coords]{{\includegraphics[width=.3\linewidth, height=.6\linewidth]
{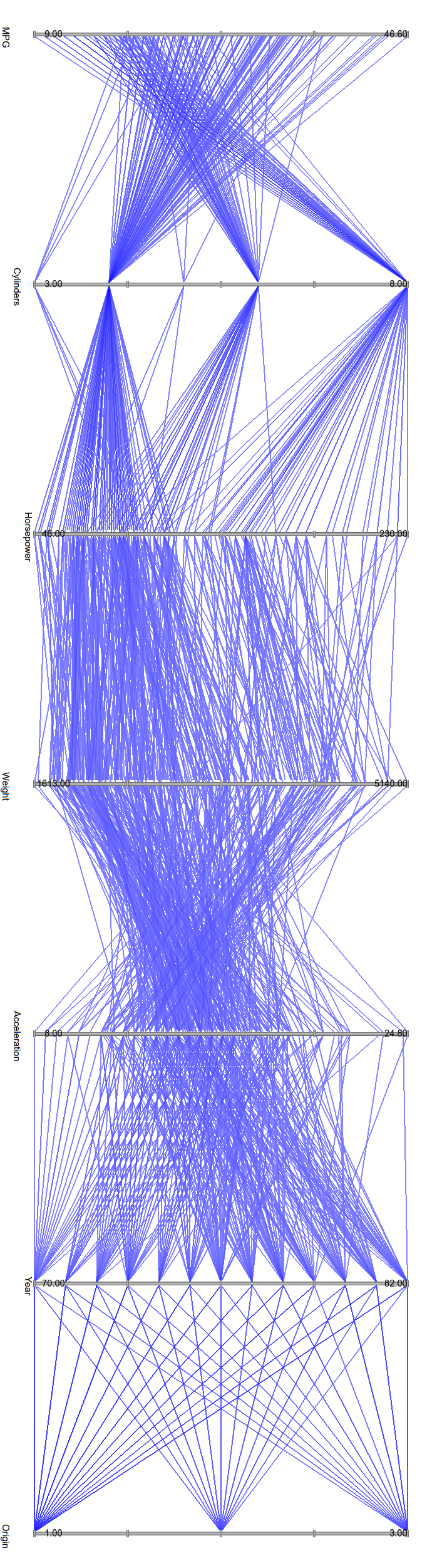}}        \label{fig-car-par}}
\hbox{
\subfigure[PCA]              {{\includegraphics[width=.24\linewidth, height=.24\linewidth]
{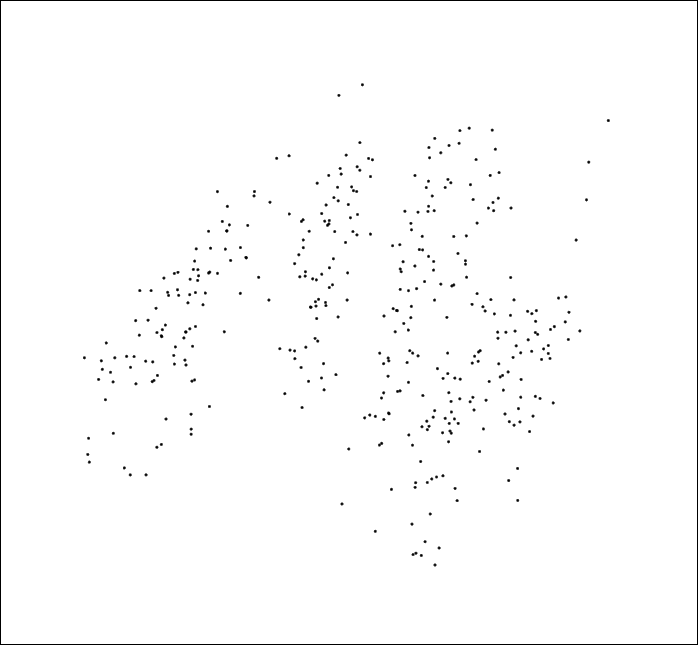}}        \label{fig-car-pca}}
\subfigure[MPG]              {{\includegraphics[width=.24\linewidth, height=.24\linewidth]
{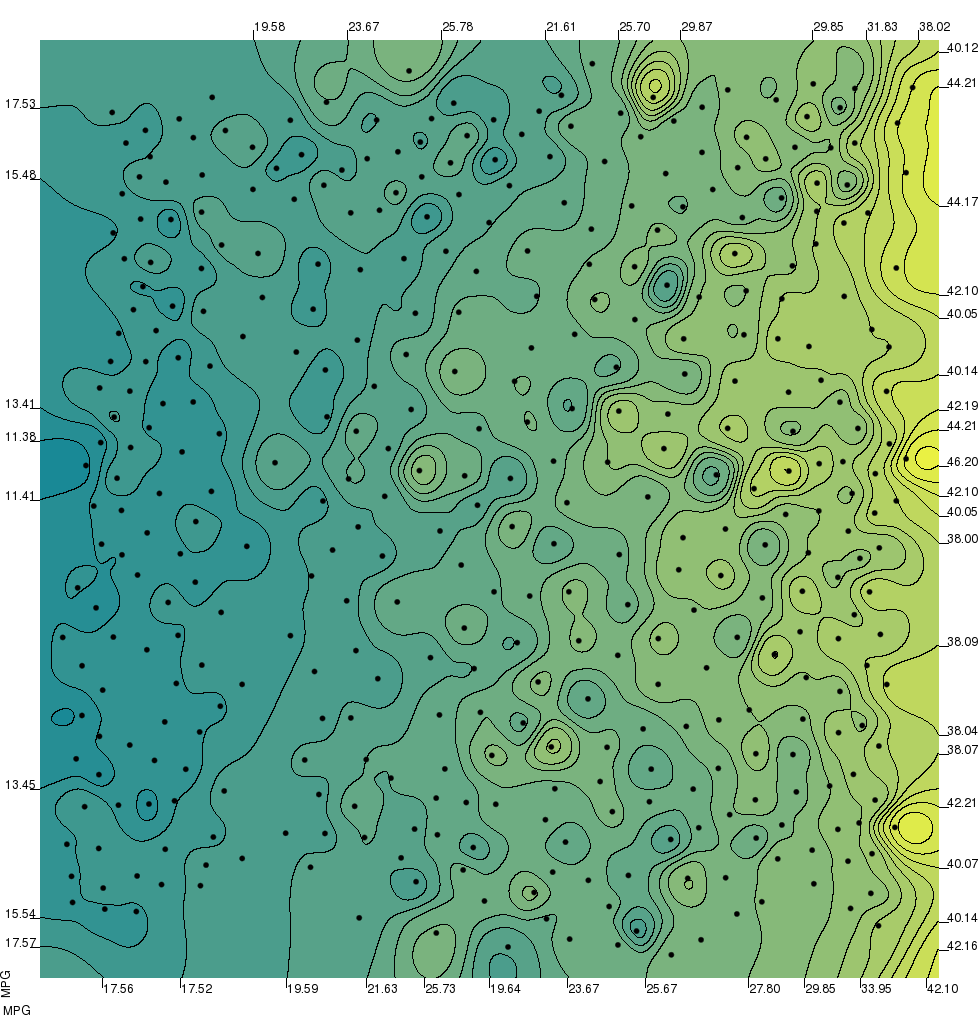}}        \label{fig-car-0}}
\subfigure[Cylinders]       {{\includegraphics[width=.24\linewidth, height=.24\linewidth]
{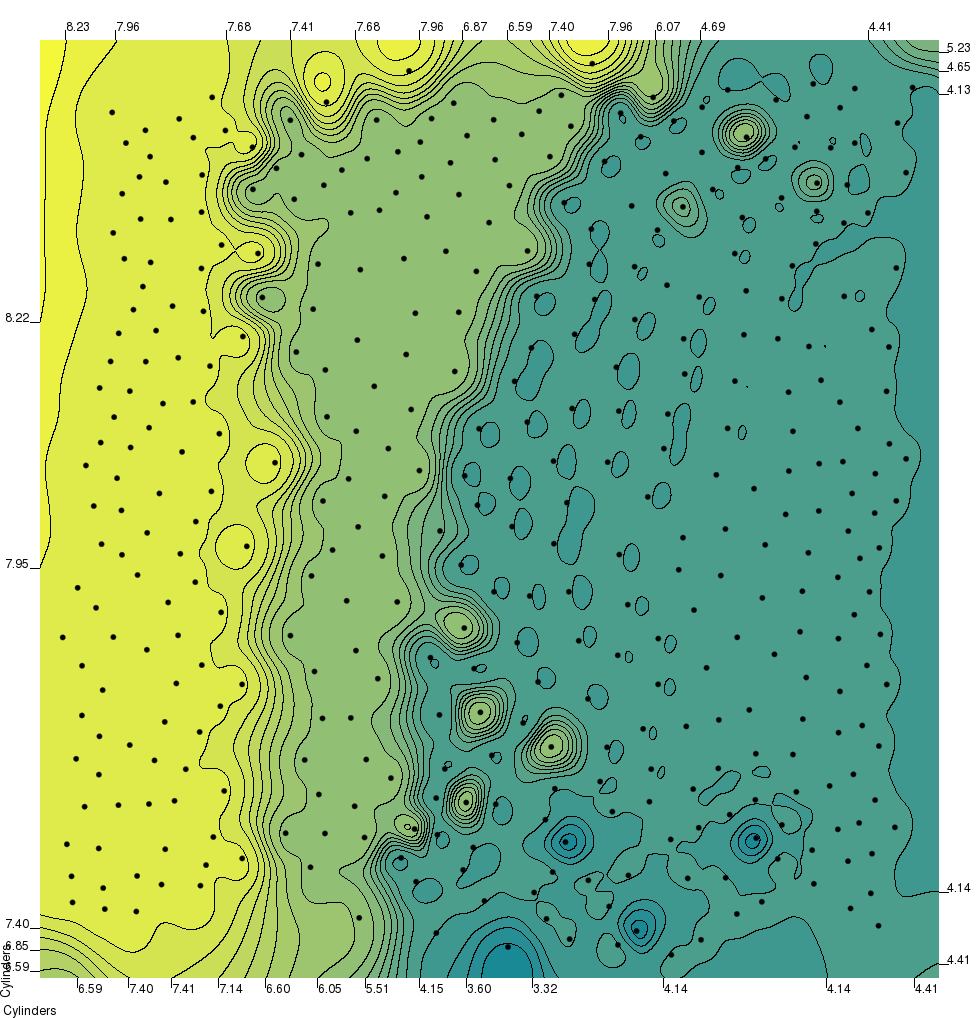}}        \label{fig-car-1}}
\subfigure[Horsepower]   {{\includegraphics[width=.24\linewidth, height=.24\linewidth]
{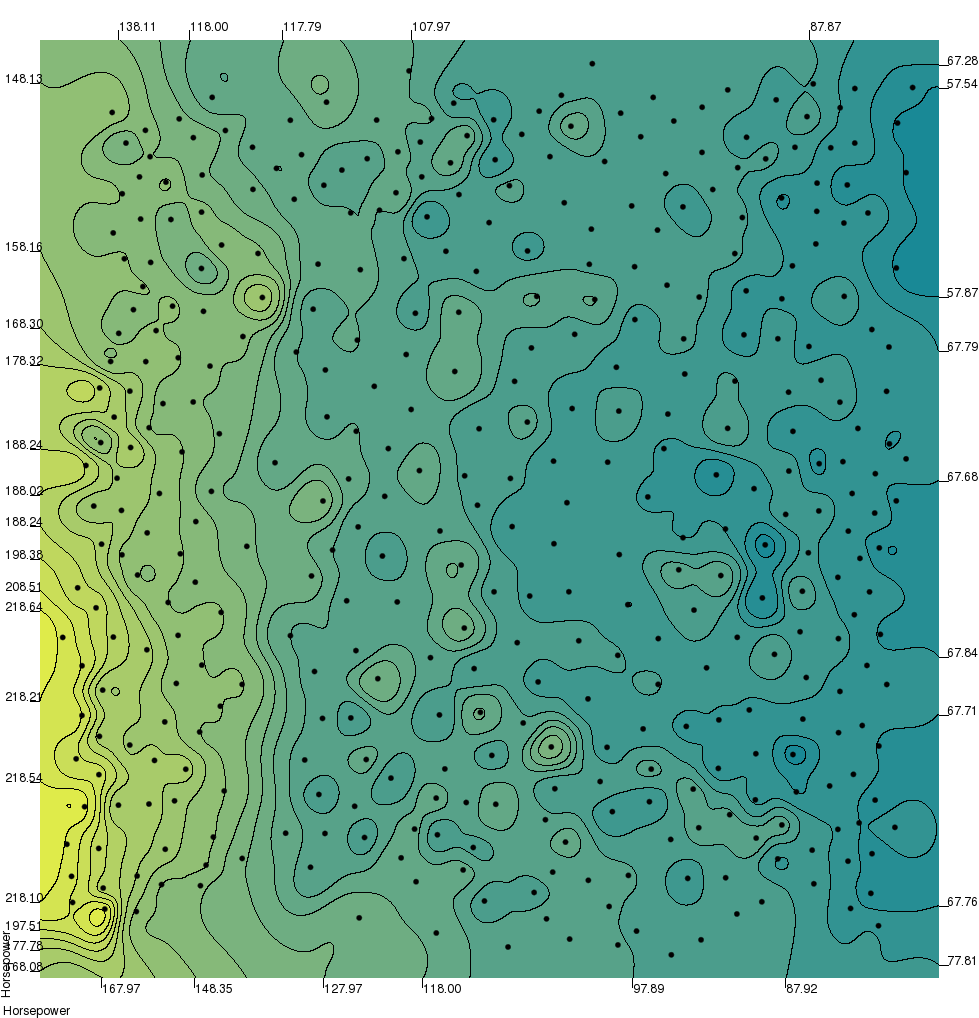}}        \label{fig-car-2}}
} \hbox{
\subfigure[Weight]           {{\includegraphics[width=.24\linewidth, height=.24\linewidth]
{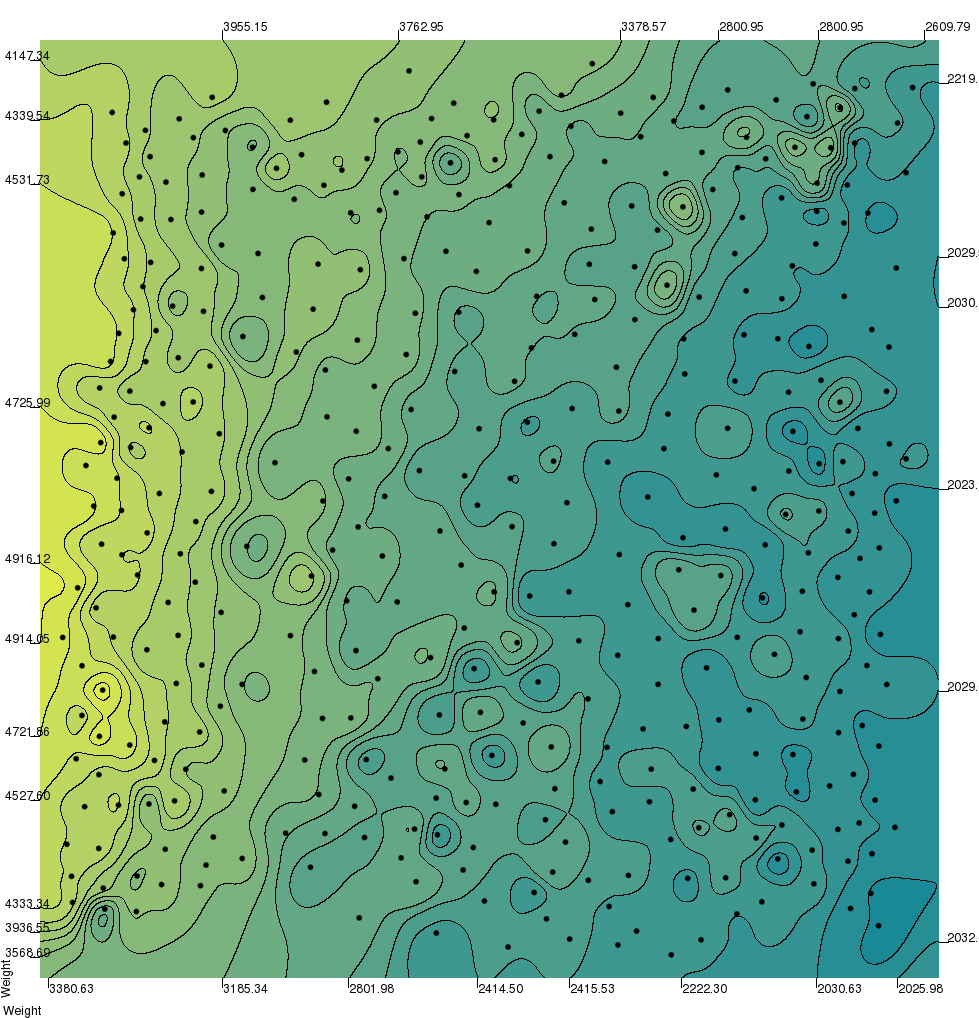}}        \label{fig-car-3}}
\subfigure[Acceleration]  {{\includegraphics[width=.24\linewidth, height=.24\linewidth]
{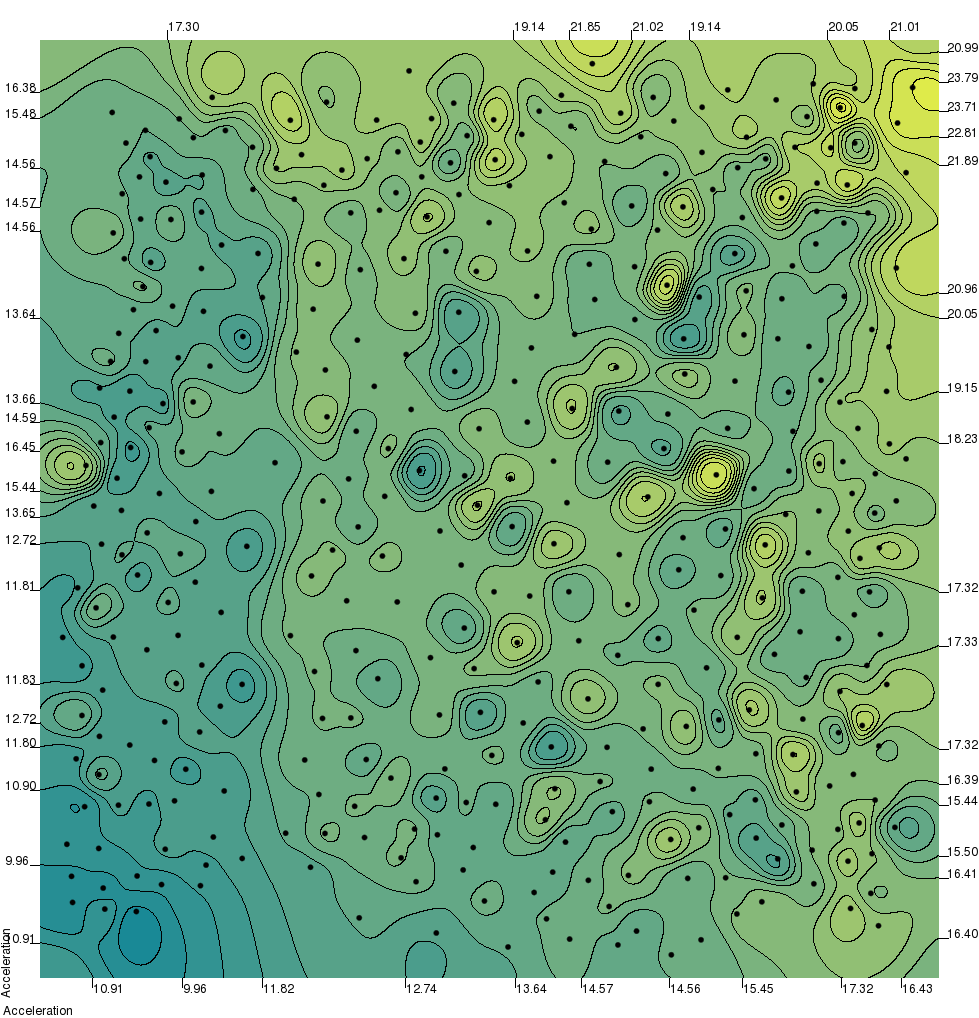}}        \label{fig-car-4}}
\subfigure[Year]              {{\includegraphics[width=.24\linewidth, height=.24\linewidth]
{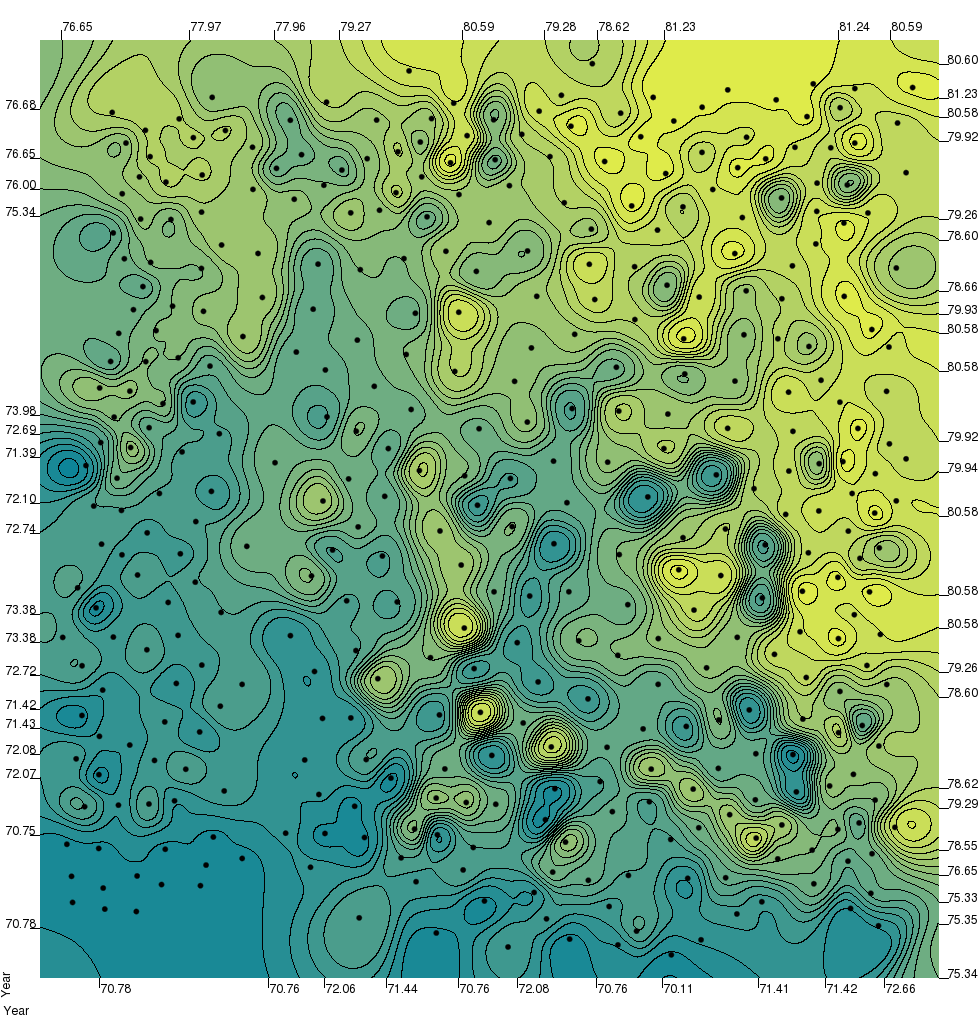}}        \label{fig-car-5}}
\subfigure[Origin]           {{\includegraphics[width=.24\linewidth, height=.24\linewidth]
{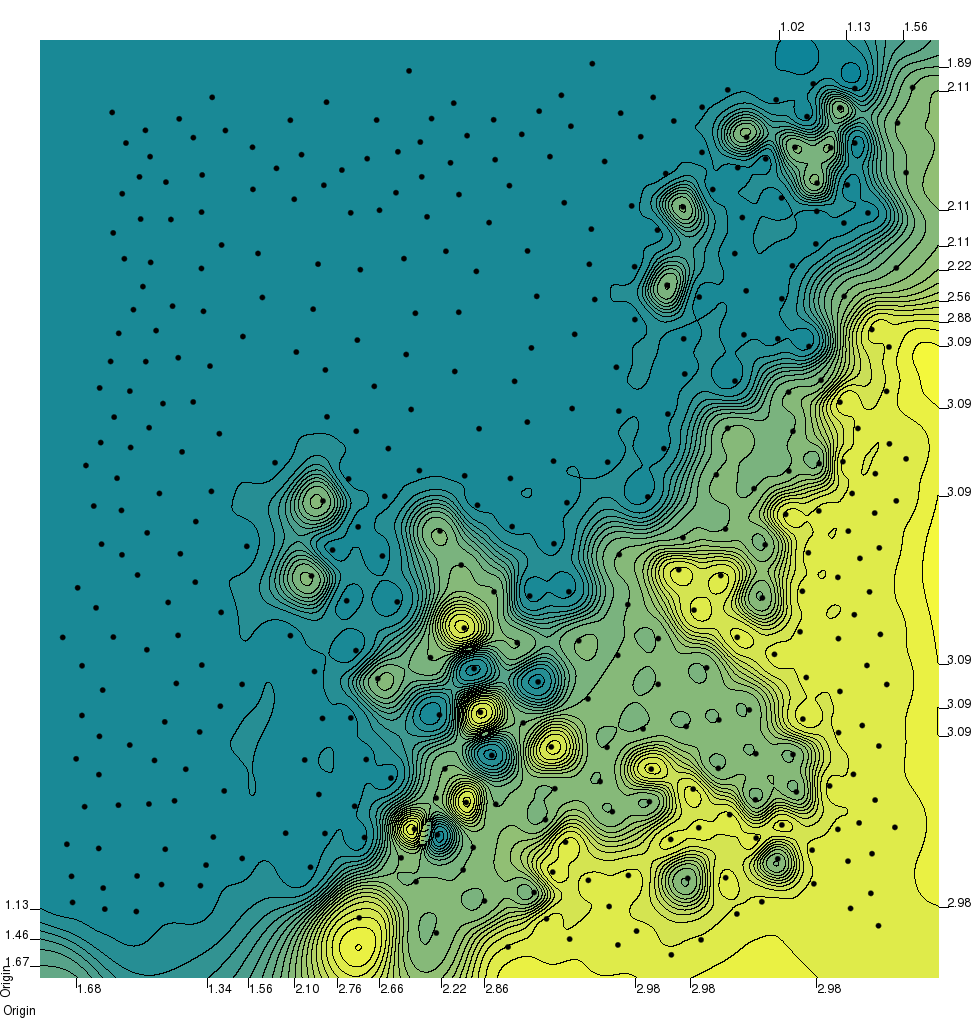}}        \label{fig-car-6}}
}
\caption{\emph{Comparison of multidimensional techniques.}  The scatterplot matrix (a) takes $O(d^2)$ views to show all dimensions spatially.  Parallel coordinates only shows $O(d)$ pair-wise dimension correlations.  In our approach, all individual dimensions are plotted separately, and can be directly compared against each other since the points don't move.  Outiers in each dimension become obvious, as are trends.}
\label{fig-cars}
\end{figure*}

\section{Evaluation} %
Similar to scatterplots, our approach visualizes a point cloud with a set of vertices on the screen.
However, it embodies a paradigm shift away from traditional scatterplots, as absolute spatial location is co-opted, and valuation of points is determined by the distorted spatial representation instead of the points' locations themselves.
For this representation to be useful, it is imperative that it convey the same information as traditional plots -- thus, its readability must be considered.  
As such. we evaluated it's readabilty with respect to several common questions that existing techniqes both excel at and perform poorly at.
More specifically, we evaluated  cluster-based questions, which projection-based methods oftem excel at but which dimension-based methods (scatterplot matrices, parallel coordinates) are known to perform poorly, 
as well as trend-based questions, which dimension-based techniques are known to be effective yet projection-based approaches often obfuscate.
In other words, we evaluated whether our approach simultaneously provided insights for which traditional representations only provide one.
We apply our method to several datasets, including several collected from IBM's ManyEyes website \cite{manyeyes} as well as some traditional multidimensional data sets such as the car and wine datasets. 
We contrast our method to existing distortion techniques and demonstrate an example where it outperforms existing methods.
Lastly, our distorted plots are nontrivial to generate, so we evaluate the performance and show that our GPU implementation is capable of interactive frame rates.

\subsection{Isocontour Patterns}
\begin{figure}
\centering
\subfigure[Peak]  {{\includegraphics[width=.45\linewidth, height=.45\linewidth]{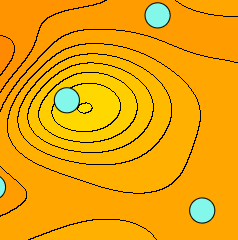}}\label{fig-peak}}
\subfigure[Valley]{{\includegraphics[width=.45\linewidth, height=.45\linewidth]{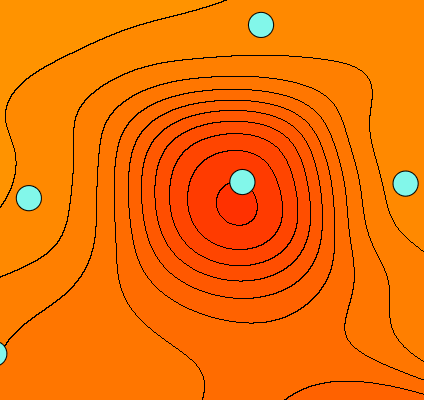}}    \label{fig-valley}}
\subfigure[Plateau]{{\includegraphics[width=.45\linewidth, height=.45\linewidth]{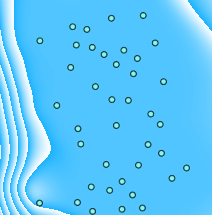}}    \label{fig-plateau}}
\subfigure[Slope]{{\includegraphics[width=.45\linewidth, height=.45\linewidth]{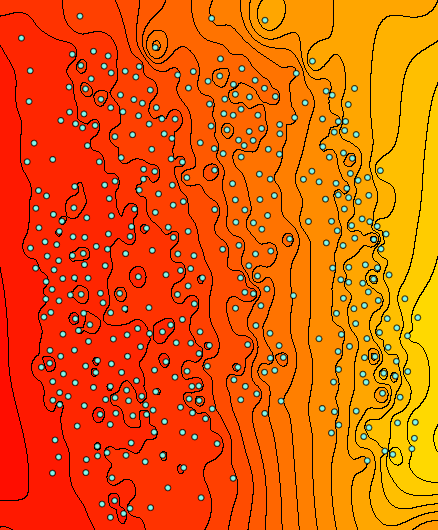}}\label{fig-slope}}
\caption{\emph{Example patterns} The interpolated and rendered space can reveal features of the dataset.  Outliers can create peaks (a) or valleys (b) when no nearby points are similar.  Plateaus (c) are indicative of data points that all have similar values (e.g. clusters or discrete dimensions).  Gradual slopes (d) are indicative of data dimensions that correspond well with the projection.  
}
\label{fig-patterns}
\end{figure}

Isocontour plots are used particularly frequently in geographic applications such as topographic maps or weather maps.  As such, many common patterns from such existing maps are relevant here.  Peaks (Figure~\ref{fig-peak}) or valleys (Figure~\ref{fig-valley}) are represented by concentric rings, and indicate local maxima or minima respectively.  When there is one data point in the middle of the rings, this indicates that that point is an outlier, as it is dissimilar from all of its neighbors.  Conversely, clusters of similar or even identical values will create plateaus (Figure~\ref{fig-plateau}), with little or no isocontours.  This way, even if clusters run up against each other due to the force directed layout, the clusters can still be easily perceived.  In the case of the gradient renderings, the color will even be constant (or nearly constant).
Just like a gradual slope in a topographic map, a series of isocontours indicates a smooth (albeit often noisy) trend in the data dimension (Figure~\ref{fig-slope}).  The distance between the isocontours, or number/darkness of contours in the adaptive case, indicate the grade of the slope, which will often roughly correspond to the number of data points in that range.

\subsection{Comparison of Multidimensional Methods}
Scatterplot matrices, parallel coordinates, and dimensional reduction methods such as PCA are all common techniques for multidimensional data analysis.  Figure~\ref{fig-cars} shows a comparison of these three approaches with our approach as applied to the cars dataset.  The scatterplot matrix (Figure~\ref{fig-car-mat}) shows pairwise comparisons between each of the 7 dimensions, for a total of $7^2=49$ images.  As there are so many scatterplots, each one becomes rather small unless paired with additional views.  There is also a lot of overplot - particularly in discrete dimensions such as cylinders or origin, which are nearly unreadable.  The parallel coordinates representation (Figure~\ref{fig-car-par}) also shows pairwise comparisons between dimensions, but only the relationship between dimensions that are next to each other is clear.  For example, it is difficult to see the relationship between MPG and weight.  Also, overplot is again an issue, particularly in the discrete dimensions.  A PCA projection (Figure~\ref{fig-car-pca}) gives an overall summary of the data, and reveals that it clusters fairly well, but all information about the source dimensions are lost.  Our appoach (Figures~\ref{fig-car-0}-\ref{fig-car-6}) reintroduces the dimensions to this projection.  Since the points are fixed between each view, any two of them are visually comparable.  This makes trends such as positive correlation between dimensions immediately obvious, as the two images will look similar, such as Horsepower and Weight (Figures \ref{fig-car-2} and \ref{fig-car-3}).  Due to the directions of the gradient in these images, it is also apparent that the Cylinders dimension (Figure \ref{fig-car-1}) also has a fairly strong correlation with these two.  Conversely, since Acceleration and MPG have gradients in the opposite direction, they have a fairly negative correlation with the other three.

\subsection{An Example Case}
In order to demonstrate the versatility of our approach, we apply it to some real data.  
Figure~\ref{teas} shows some examples from the wine dataset.  While the traditional PCA plot (Figure~\ref{teas1}) shows one general trend, the underlying data is much more complex.   The flavonoid dimension (Figure~\ref{teas2}) reveals that it pretty strongly correlates with the first principal component, with the exception of one outlier in the middle of the plot and another near the bottom left.  
The alcohol dimension (Figure~\ref{teas3}) is mostly orthogonal, to this dimension, and reveals other outliers.

\subsection{Pilot User Study}

In order to evaluate the usefulness and effectiveness of our representation, we performed a comparative user study.
In this study, we presented users with a series of questions with randomly selected, relevant representations to measure the difference in their performance for a number of simple tasks pertaining to point evaluation, trend evaluation, and cluster perception.  These tasks were selected such that many existing techniques would excel at one or two tasks, but not all, and we hypothesized that our approach would handle all tasks well.  Specifically, we asked the users:
\begin{itemize}
\item What is the value of a specific data point?
\item How many clusters are there in the data?
\item How many points are in a specific cluster?
\item Which dimensions are positively/negatively correlated?
\end{itemize}

Not every question makes sense for every representation.  For instance, there is little value to asking point valuation questions on a dimensional projection.  
However, it is notable that our representation is viable for all of these questions, whereas the existing approaches might only be effective for a subset.
So we selected representations for each question that helped to comparatively evaluate aspects of the particular question with techniques known to be effective for the task, 
while skipping existing representations where the stated question would be impossible (e.g. asking for point valuation on a projected plot).
The valuation question in particular was asked twice: once to compare our approach against traditional approaches, and once to compare the different variations of our approach.  

There were about 40 users, ranging in age from 22 to 65, and of all levels of expertise in visualization and data analysis.

The results of this pilot study confirmed several of our hypotheses.  Firstly, we did confirm that ou representation was viable for each of the stated tasks, whereas traditional representations are only applicable to some of them.
One preliminary conclusion from this user study is that users are more accurate in estimating the value of a point when an MLS-based interpolation method is used, rather than simple linear interpolation. Users were also better at estimating the size of clusters when shown contour plots as opposed to plain PCA plots, as PCA resulted in dense regions of points overlapping. Finally, our method was on par with scatterplot matrices and far better than parallel coordinates for the task of determining which dimensions exhibit correlations.

\subsection{Performance}
The computation time required for our approach to produce a contour plot has 3 parts: triangulation, layout, and MLS. Table \ref{tab-time} shows running times for each of these steps on different scales of data. These tests were performed on a Linux machine with 8x Intel Xeon X5450 3GHz,16GB of system RAM, and a GeForce GTX 285 with 2GB of GPU RAM and 240 CUDA cores, and a 600x600 canvas.

The triangulation is a pre-processing step, and bears no impact on interactivity. Also, the results of our timing tests indicate that the triangulation computation time is negligible, so there would be little benefit to further optimization such as GPU acceleration.
The time complexity of our force-directed layout is $O(|V|\log{|V|} + |E|)$ per iteration. Since Delaunay triangulation guarantees a constant number of triangles per vertex on average, the bound reduces to $O(|V|\log{|V|})$. Planarity constraints add a linear scaling factor (albeit a small one) to the per-iteration complexity, in addition to increasing the convergence time. Thus, the layout is the most time consuming step of the process. However, with GPU acceleration this step can still be performed reasonably quickly. Also, the layout only needs to be recomputed when its parameters are adjusted and the user explicitly recalculates the layout, so it has no impact on interactivity. 

Since the MLS mapping is the last step of the process, it is recomputed when any parameter is adjusted. The MLS computation scales with the resolution and the number of points in the plot, giving it a time complexity of $O(w h n)$. But each pixel can be calculated independently, making it optimal to apply GPU acceleration. Monitor resolution limits present more of a upper bound on plot size than onerous computation time.  Turning off MLS and using the standard linear interpolation method is fastest, as it has $O(1)$ computation complexity per pixel. While it does not look as nice as the other modes, its speed can make it useful for adjusting settings such as layout parameters.  Each of the MLS modes perform slightly differently, but are quite comparable.  The mean MLS calculation was the fastest of the MLS methods, performing nearly twice as fast as the other two, with results nearly equivalent to rigid MLS.  However, the affine MLS results were the smoothest and nicest looking, making it usually worth the extra computation. Notably, all renders were at interactive frame rates, ranging from as good as over 30 FPS down to about 4 FPS.

\begin{table}
\centering
\caption{Timing tests. Duration values are in seconds.}
\label{tab-time}
\begin{tabular}{r|r|r|r|}
\# Nodes: & 306 & 1000 & 2245\\
\hline
Triangulation: & 0.000607 & 0.002450 & 0.004036\\
Layout: & 0.783082 & 1.314509 & 1.777599\\
\hline
Linear Interp.: & 0.000631 & 0.001076 & 0.001363 \\
Mean MLS:   & 0.017420 & 0.064938 & 0.147089 \\
Affine MLS:   & 0.026670 & 0.100724 & 0.242192 \\
Rigid MLS:   & 0.028846 & 0.116083 & 0.246789 \\
\end{tabular}
\end{table}

\subsection {Limitations}
While MLS does have guarantees, such as smoothness and nice interpolation, it is not always a 1-to-1 mapping. This can be seen in the examples of `foldback' in previous works \cite{imagemls}. In our approach, we avoid explicit `foldback' by preserving the planarity of the triangular mesh.  However, similar artifacts can show up as local extrema, which are rendered as closed loops with no contained data points. 
These extrema can often be mitigated, though not always completely eliminated, by exploratively adjusting the MLS exponent $\alpha$ in equation \ref{eq-weights}. When $\alpha$ is high, the isocontours that surround points are reduced, but the frequency and amount of stand-alone contours increases and the smoothness of the plot can suffer. When the exponent is low, it reduces the occurrence of extrema between points, but increases the number of isocontours that surround individual points. If the exponent is too low ($\alpha \approx 0$), these isocontours actually shrink to a subpixel level and disappear, which will make the plot visually inaccurate. Thus, there is a tradeoff between low and high $\alpha$ values which generally requires user interaction. It is noteworthy that the affine MLS suffers much less from stand-alone contours, and thus can handle much higher $\alpha$ values than the other variants.

There is a performance limitaiton to scaling our approach, since it performs per-pixel operations with $O(n)$ computations.  This could be reduced by sampling screen-space at a lower resolution, or by moving away from MLS and to a more localized sampling algorithm.  However, every step of our approach is easily parallel, so GPU acceleration alleviates this.

In our initial exploration, we also investigated simultaneously showing more than 1 dimension with isocontours.  While algorithmically sound, we found that with few exceptions, the additional visual complexity of showing 2 or more dimensions substantially detracted from legibility.  Thus, we did not even consider it in our user evaluation.

\section{Future Work} %
There are several optimizations that could be implemented to improve the efficiency of the approach. As in \cite{imagemls}, it could be beneficial to pre-calculate portions of the MLS equations that are independent of the distorted points, to allow for more interaction with the points. While the layout we use is GPU accelerated, many of the multi-level optimizations of FM$^3$ had to be limited or removed due to the planarity constraints. We currently use a CPU implementation of the Delaunay triangulation, but there do exist GPU accelerated versions that could reduce the time to create an initial view.
One of the limitations of MLS is that, even with the planarity guarantees, it is still possible for space to get warped in such a way that it folds on itself, causing local maxima and minima. Alternative interpolation based on high-dimensional splines, a line variant of MLS \cite{imagemls}, a variant of Sibson's interpolant \cite{sibson}, or some other coordinate interpolation algorithm such as \cite{mean-coordinates} might address this issue.
Another limitation to MLS is that every pixel depends on every data point, not just the nearest ones, meaning that our approach would not scale well to very large datasets.  Sibson's interpolant, particularly a GPU accelerated approximation \cite{sibson}, should scale better than MLS to very large data sets.
It would be beneficial to choose data-dependent, semantically meaningful spacings for the lines on each axis, as in \cite{2010-tick-labels}.
When dealing with very high dimensional data, it could also be helpful to organize the plots to place similar dimensions next to each other for ease of comparison, in a self organizing map for instance \cite{som-org}.  
Finally, while we designed these plots with readability in mind, and our cursory user study confirmed several of our intended hypotheses, 
a more comprehensive user study would help to rigorously evaluate our approach's perceptual effectiveness.

\section{Conclusion} %
Multidimensional data analysis is among the most common of information visualization tasks, and our contour based method is a novel approach which incorporates and improves methods from a variety of domains, including dimensional reduction, graph drawing, and iso-contouring.  
While each of these disciplines has a long history, we have shown how their combination can yield useful results.  
Even some of the individual components such as the graph layout are practical on their own and could be applied to many other problems.

\bibliographystyle{abbrv}
\bibliography{paper}
\end{document}